\documentclass[fleqn,usenatbib]{mnras}
\usepackage{newtxtext,newtxmath}
\usepackage[T1]{fontenc}
\usepackage{ae,aecompl}
\usepackage{graphicx} 
\usepackage{hyperref}
\DeclareUnicodeCharacter{03BB}{\ensuremath{\lambda}} % lowercase lambda
\DeclareUnicodeCharacter{039B}{\ensuremath{\Lambda}} % capital lambda
\title[Quarkyonic Matter] {Quarkyonic Model for Neutron Star Matter: A Relativistic Mean-Field Approach}
\author[Kumar et al.]{
Ankit Kumar$^{1,2}$
\thanks{ankit.k@iopb.res.in},
Debabrata Dey$^{1,2}$, Shamim Haque$^{3}$, Ritam Mallick$^{3}$, S. K. Patra$^{1,2}$
\\
$^{1}$ Institute of Physics, Sachivalaya Marg, Bhubaneswar-751005, India\\
$^{2}$ Homi Bhabha National Institute, Training School Complex, 
Anushakti Nagar, Mumbai 400094, India\\
$^{3}$ Indian Institute of Science Education and Research Bhopal, Bhopal 462066, India}
\begin{document}
\maketitle
%\linenumbers
%\date{\today}
%%%%%%%%%%%%%%%
\begin{abstract}
The concept of quarkyonic matter presents a promising alternative to the conventional models used to describe high-density matter and  provides a more nuanced and detailed understanding of the properties of matter under extreme conditions that exist in astrophysical bodies. The aim of this study is to showcase the effectiveness of utilizing the quarkyonic model, in combination with the relativistic mean-field formalism, to parameterize the equation of state at high densities. Through this approach, we intend to investigate and gain insights into various fundamental properties of a static neutron star, such as its compositional ingredients, speed of sound, mass-radius profile, and tidal deformability. The obtained results revealed that the quarkyonic matter equation of state (EOS) is capable of producing a heavy neutron star with the mass range of $\sim$ $2.8 M_\odot$. The results of our inquiry have demonstrated that the EOS for quarkyonic matter not only yields a neutron star with a significantly high mass but also showcases a remarkable degree of coherence with the conformal limit of the speed of sound originating from deconfined QCD matter. Furthermore, we have observed that the tidal deformability of the neutron star, corresponding to the EOSs of quarkyonic matter, is in excellent agreement with the observational constraints derived from the GW170817 and GW190425 events. This finding implies that the quarkyonic model is capable of forecasting the behaviour of neutron stars associated with binary merger systems. This aspect has been meticulously scrutinized in terms of merger time, gravitational wave signatures and collapse times using numerical relativity simulations.
\end{abstract}
%%%%%%%%%%%%%%%%
%%%%%%%%%%%%%%
%%%%%%%%%%%%%%%%%%%%%%%%%%%%%%%%%%%%%%%%%%%%%%%%%%%%%%%%%%%%%%%%%%%%%%%%%
\section{Introduction}\label{intro}
Neutron stars are incredibly intriguing astronomical objects that arise from the aftermath of a supernova explosion, during which the star's core becomes exceedingly dense, forming a tightly packed mass of neutrons with distinctive properties such as a diameter of only around $10-15$ kilometers and a mass that can be $1.1$ to $2.5$ times that of our sun \citep{doi:10.1146/annurev-nucl-102711-095018}. Despite their diminutive size, neutron stars are incredibly potent objects that play a pivotal role in the cosmos, and the study of them helps us to better comprehend the physics of extreme environments and unravel the enigmas of the universe. The investigation of neutron stars requires a diverse range of scientific disciplines, including general relativity, high-energy physics, nuclear and hadronic physics, neutrino physics, quantum chromodynamics, superfluid hydrodynamics, plasma physics, and even solid-state physics. This interdisciplinary approach provides a comprehensive understanding of neutron stars and enables us to scrutinize the behavior of matter in the most extreme environments. 

Neutron stars are also of great interest to astronomers because they emit copious amounts of radiation, including X-rays and gamma-rays, making them detectable across the electromagnetic spectrum. Observational data from pulsars, rapidly rotating neutron stars with intense magnetic fields, have been instrumental in our understanding of neutron star properties. Recent breakthroughs in observation and theory, including the discovery of kHz quasi-periodic oscillations, bursting millisecond pulsars, and half-day-long X-ray superbursts \citep{2002A&A...382..503K, 2003HEAD....7.1702W, Linares_2005, Keek_2012, 10.1093/mnras/stt919, 2019A&A...621A..53I, Bult_2019, 2021cosp...43E1213G, 2022A&A...658A..95V, 2022A&A...664A..54G}, have provided us with an enormous abundance of observational data, empowering us to test and refine our theoretical models, leading to novel insights into the properties of matter and the laws of physics in the most extreme environments. Additionally, the thermal emission from solitary neutron stars permits the measurement of their radii and offers crucial information regarding their cooling history. At the same time, advances in radio telescopes and interferometric methods have increased the number of known binary pulsars \citep{Tyulbashev2016, 10.1093/mnras/stab790}, allowing for tests of general relativity and incredibly accurate measurements of neutron star masses and offering the possibility of discovering new physics beyond our current understanding.

Neutron star research is a rapidly growing field in astrophysics, as these objects provide a unique opportunity to study matter under extreme conditions. Of particular interest is the equation of state (EOS) of neutron star matter, which relates the system's energy and pressure to its temperature, density, and composition, ultimately determining macroscopic properties such as mass, radius, and moment of inertia. Obtaining the EOS can help provide a comprehensive understanding of nuclear matter properties at all densities, given a proper parameter set that satisfies various observational nuclear matter constraints. The EOS of dense matter at both sub-nuclear and supranuclear densities is a challenging task, and research has been focused on using laboratory experiments involving heavy ion collisions, supernova simulations, and theoretical many-body formalism to study the EOS of neutron star matter \citep{PhysRevLett.120.172703, foundations1020017, doi:10.1146/annurev-nucl-102419-124827, Russotto2023}. Despite the difficulties associated with making accurate predictions about the EOS due to the extreme conditions found in neutron stars, recent progress has been made through a combination of theoretical and observational advancements. The observation of gravitational waves from neutron star mergers provides a new way to study these objects and probe matter's properties under the most extreme conditions. Additionally, advances in computational power and numerical techniques enable more accurate and detailed simulations of neutron star mergers, providing important insights into the EOS.

The relativistic-mean field (RMF) model is a highly effective and extensively used theoretical framework to challenge a longstanding puzzle of EOS of extremely dense astrophysical matter and for investigating a range of phenomena, including neutron stars, infinite nuclear matter, and finite nuclei. This approach employs the concept of self-consistency, which is used to determine the equations of motion for the nucleons, taking into account their interactions with mesons \citep{WALECKA1974491, BOGUTA1977413, Serot:1984ey, SEROT1979146}. The RMF model has been successful in explaining various properties of symmetric and asymmetric nuclear matter, including nuclear binding energies, density distributions, and ground-state properties of nuclei. It has also been applied to explore the astrophysical compact objects, where it has been used to derive the EOS and investigate their structural properties \citep{LALAZISSIS19991, PhysRevC.63.044303, PhysRevLett.86.5647, PhysRevLett.95.122501, PhysRevC.82.055803, 10.1093/mnras/stab2387}. Additionally, it has also been used to investigate the effect of different interactions, such as hyperons, on the properties of neutron stars, providing valuable insight into the nature of dense matter \citep{1985ApJ...293..470G, PhysRevC.52.3470, PhysRevC.85.065802, PhysRevC.97.015805, Biswal_2019, PhysRevD.104.123006}. Due to its success and versatility, the RMF model continues to be a valuable tool for understanding the behavior of matter under extreme conditions.

The major constituents of a neutron star are protons and neutrons, which are basically bound states of quarks. When the nuclear matter becomes so dense that the hadrons start to overlap, and the quarks in different hadrons can be exchanged, it is most suitable to describe the nuclear dense matter with quark degrees of freedom. As a consequence, a phase transition takes place between nuclear matter and quark matter, which is a common occurrence within the core of a neutron star as well as in heavy ion collisions. Incorporating quark matter into current theoretical models to study the observables of neutron stars has had a significant impact on the maximum mass of a neutron star, as well as its radius and cooling behavior.

Quarkyonic matter in \cite{PhysRevLett.122.122701} is a novel form of matter that is distinct from both quark-gluon plasma and nuclear matter. It is characterized by a dense assembly of quarks and gluons that are confined within a finite region, giving rise to a new state of matter that is neither purely hadronic nor purely quark-gluon plasma. The quarkyonic matter is expected to occur at intermediate densities, where the number of quarks and gluons is not large enough to form a plasma but is still large enough to interact strongly. Quarkyonic matter is peculiar in the way that the speed of sound within it does not follow a consistent pattern as the matter's density increases. Rather, it demonstrates a distinct trend where the sound velocity first attains its maximum value at a lower density, subsequently declines, and then rises again until it reaches a maximum of $1/\sqrt{3}$ \citep{PhysRevLett.122.122701}. Typically, the addition of extra degrees of freedom, such as pions, hyperons, dark matter particles, etc., to proto-neutron star matter typically results in a decrease in neutron star mass, however, quarkyonic matter sets itself apart by supporting neutron stars with larger radii and greater maximum mass, which is in contrast to previous models. Quarkyonic matter possesses unique properties that enable it to generate higher pressure across a range of energy densities within the core of a compact star. In addition, the quarkyonic matter may provide a consistent explanation for several phenomena such as the saturation of the nuclear matter equation of state at high densities, the suppression of high transverse momentum hadrons in heavy-ion collisions and various astrophysical observations, including the damping of r-mode oscillations in all millisecond pulsars and the low temperatures observed in low-mass X-ray binaries. These characteristics makes quarkyonic matter an exciting area of research with the potential to yield new insights and discoveries in astrophysics. 

The Quarkyonic model put forth by McLerran and Reddy is a simplified rendition that focuses on a solitary nucleon species and a two-flavored quark system that is unburdened by charges of u and d quarks. However, the model fails to account for the necessities of chemical or beta equilibrium, which entails the minimization of energy concerning particle and charge densities at all levels of density. Furthermore, protons and leptons are missing from the model, thereby rendering it incapable of achieving chemical or beta equilibrium. Later, several groups attempted to improve the quarkyonic model by introducing an excluded volume hard core potential for nucleons, thereby enabling it to achieve quark-nucleon chemical equilibrium and integrate beta equilibrium \citep{PhysRevC.101.035201, PhysRevC.102.025203, universe8050264}. In a separate publication, Zhao et al. introduced a quarkyonic model that encompasses protons and leptons and successfully fulfills the essential requirements of chemical and beta equilibrium \citep{PhysRevD.102.023021}. They propose a modified approach to the chargeless 2-flavor quark model that incorporates asymmetrical nucleon matter and leptons. Their model was able to conform to the experimental and observational limitations linked to neutron star structure. Nonetheless, the potential utilized by Zhao et al. only applies to asymmetric nuclear matter and fails to accurately describe the precise physical variations of constituents inside a neutron star. The coefficients of the potential used by the former authors in \citep{PhysRevD.102.023021} are fitted either for pure neutron matter or symmetric nuclear matter to satisfy the nuclear matter constraints at lower densities, and so the interacting potential for nucleon after transition density remains unaffected by values of the minimum allowed Fermi momentum of the quarkyonic model. In contrast, to include the effects of the minimum allowed Fermi momentum in the potential part of the energy density expression, to estimate the effective chemical potential necessary for the determination of the mass of quarks at the transition density, and for a more consistent beta-equilibrated matter's potential rather than the symmetric or pure nucleon matter, the interaction potential used in our work calculated by using the RMF formalism is much suitable, which is much different from the one used by previous authors and resolve the above-mentioned issues. We employed a theoretical mean-field approach in conjunction with the quarkyonic model to derive a more cohesive equation of state for quarkyonic-neutron star matter. To ensure beta equilibrium and charge neutrality in the presence of both nucleons and quarks, we primarily rely on the calculations conducted by Zhao et al. However, in our scenario, we determine the energy and pressure of the nucleons using the RMF formalism, while ensuring beta equilibrium for both quarks and nucleons within the same framework. Following the derivation of the EOS using this approach, we proceed to compute various properties of both static and rotating neutron stars. We then compare these results with the constraints established through the analysis of various observational gravitational wave events. In addition, we utilize the EOS obtained from our calculations to investigate the dynamics of binary star merger events.

The present paper is structured as follows. In Section 2, we undertake the task of obtaining the equation of state (EOS) of a neutron star with the quarkyonic matter by employing the RMF formalism with G3 \citep{KUMAR2017197} and FSUGold \citep{PhysRevLett.95.122501} parameter sets. This approach has been widely utilized for studying the properties of dense nuclear matter, making it a suitable choice for our investigation. Moving on to Section 3, we utilize the derived EOS to explore several properties of a static quarkyonic neutron star, such as the speed of sound, mass-radius profile, and tidal deformability. Additionally, we analyze observational data to gain a deeper understanding of these properties.  
%In Section 4, our focus shifts towards gaining insights into the Kepler frequency and spin parameter of a rotating star. 
Moving on to the next section, we discuss the inferences derived from 3D numerical relativity simulations of binary neutron star mergers (BNSM) of equal mass binaries. Finally, we present some conclusions in the last section.
%%%%%%%%%%%%%%%%%%%%%%%%%%%%%%%%%%%%%%%%%%%%%%%%%%%%%%%%%%%%%%%%%%%%%%%%%%%%
\section{Equation of State}\label{EoS}
%%%%%%%%%%%%%%%%%%%%%%%%%%%%%%%%%%%%%%%%%%%%%%%%%%%%%%%%%%%%%%%%%%%%%%%%%%%%
The RMF theory provides a phenomenological description of the nuclear many-body problem. It can be applied to model dense nuclear matter at different densities and temperatures, both inside finite nuclei and a neutron star. The standard approach involves constraining the couplings of the interacting nucleons via mesons and the self-interacting mesons by comparing its predictions for the symmetric nuclear matter at saturation density with measured finite nuclear properties. The theory is then extrapolated to neutron stars, subject to additional charge neutrality and beta equilibrium conditions along with the other astrophysical constraints  \citep{PhysRevC.45.844, 2018APS..APRS11009H} . Historically, various parameter sets have been developed to satisfy different observational and experimental constraints, each with its own advantages and disadvantages \citep{PhysRevC.55.540, PhysRevC.63.044303, PhysRevC.74.045806, PhysRevC.70.058801, LALAZISSIS200936, PhysRevC.82.055803, PhysRevC.82.025203, PhysRevC.84.054309, PhysRevC.85.024302, PhysRevC.97.045806, PhysRevC.102.065805}.
In this work, the Lagrangian density function for nucleons interacting via mesons can be expressed as \citep{Reinhard_1989, PhysRevC.70.054309, Kumar2020}
%%%%%%%%%%%%%%%%%%%%%%%%%%%%%%%%%%%%%%%%%%%%%%%%%%%%%
\begin{eqnarray}\label{lag}
{\cal L} & = &  \sum_{i=p,n} \bar\psi_{i}
\Bigg\{\gamma_{\nu}(i\partial^{\nu}-g_{\omega}\omega^{\nu}-\frac{1}{2}g_{\rho}\vec{\tau}_{i}\!\cdot\!\vec{\rho}^{\,\nu})
-(M-g_{\sigma}\sigma
\nonumber\\
&&
-g_{\delta}\vec{\tau}_{i}\!\cdot\!\vec{\delta})\Bigg\} \psi_{i}
-\frac{1}{2}m_{\sigma}^{2}\sigma^2
+\frac{1}{2}\partial^{\nu}\sigma\,\partial_{\nu}\sigma
+\frac{1}{2}m_{\omega}^{2}\omega^{\nu}\omega_{\nu}
\nonumber \\
&& 
-\frac{1}{4}F^{\alpha\beta}F_{\alpha\beta}
+\frac{1}{2}m_{\rho}^{2}\rho^{\nu}\!\cdot\!\rho_{\nu} 
-\frac{1}{4}\vec R^{\alpha\beta}\!\cdot\!\vec R_{\alpha\beta}
-\frac{1}{2}m_{\delta}^{2}\vec\delta^{\,2}
\nonumber \\
&&
+\frac{1}{2}\partial^{\nu}\vec\delta\,\partial_{\nu}\vec\delta
-g_{\sigma}\frac{m_{\sigma}^2}{M}\Bigg(\frac{\kappa_3}{3!}+\frac{\kappa_4}{4!}\frac{g_{\sigma}}{M}\sigma\Bigg)\sigma^3
\nonumber\\
 &&
+\frac{1}{2}\frac{g_{\sigma}\sigma}{M}\Bigg(\eta_1+\frac{\eta_2}{2} \frac{g_{\sigma}\sigma}{M}\Bigg)m_\omega^2\omega^{\nu}\omega_{\nu}
+\frac{\zeta_0}{4!}g_\omega^2(\omega^{\nu}\omega_{\nu})^2 
\nonumber\\
 &&
+\frac{1}{2}\eta_{\rho}\frac{m_{\rho}^2}{M}g_{\sigma}\sigma(\vec\rho^{\,\nu}\!\cdot\!\vec\rho_{\nu})
-\Lambda_{\omega}g_{\omega}^2g_{\rho}^2(\omega^{\nu}\omega_{\nu})(\vec\rho^{\,\nu}\!\cdot\!\vec\rho_{\nu})
\nonumber\\
 &&
+\sum_{j=e^{-},\mu} \bar\phi_{j}(i\gamma_{\nu}\partial^{\nu}-m_{j})\phi_{j}
\label{eq1}
\end{eqnarray}
%%%%%%%%%%%%%%%%%%%%%%%%%%%%%%%%%%%%%%%%%%%%%%%%%%%%%%%%%%%%%%%%%%%%%%%%%%%%
The wave functions of the nucleons (protons and neutrons) are represented by $\psi_{i}$ and the last term of the expression stands for the non-interacting leptonic part i.e. electrons and muons. The mass of the nucleon is denoted by M ($\approx939 MeV$), while the masses and coupling constants for the sigma ($m_\sigma$, $g_\sigma$, $\kappa_3$, $\kappa_4$,), omega ($m_\omega$, $g_\omega$, $\zeta_0$, $\eta_1$, $\eta_2$), rho ($m_\rho$, $g_\rho$, $\eta_\rho$, $\Lambda_\omega$), and delta ($m_\delta$, $g_\delta$) mesons are denoted separately. The field strength tensors $F^{\alpha\beta}$ and $\vec R^{\alpha\beta}$ are used for the omega and rho mesons, respectively.  In order to perform further calculations, we employ the relativistic mean-field approximation, wherein the meson fields are replaced with their average values. This simplifies the calculation process, particularly in the case of uniform static matter, where spatial and temporal derivatives for mesons can be safely ignored. The translational and rotational invariance as well as the isotropy of nuclear matter also has a bearing on the calculation, as only the time-like components of the isovector field and the isospin 3 component of the mesonic field are significant \citep{Gambhir1989, doi:10.1142/S0218301397000299, KUBIS1997191, PhysRevC.65.045201}. This ensures that the calculation remains consistent and accurate. The equation of motion for the nucleon in the RMF approximation is described by the Dirac equation, which is utilized to study the behavior and properties of nucleons in the system. 
%%%%%%%%%%%%%%%%%%%%%%%%%%%%%%%%%%%%%%%%%%%%%%%%%%%%%%%%%%
\begin{eqnarray}
\Bigg\{i\gamma_{\nu}\partial^{\nu}-g_{\omega}\gamma_{0}\omega-\frac{1}{2}g_{\rho}\gamma_{0}\tau_{3i}\rho-(M_i-g_{\sigma}\sigma- \tau_{3i}g_{\delta}\delta)\Bigg\} \psi_i=0 \nonumber,
\end{eqnarray}
%%%%%%%%%%%%%%%%%%%%%%%%%%%%%%%%%%%%%%%%%%%%%%%%%%%%%%%%%%
The equation of motions for the sigma, omega, rho, and delta mesons in the RMF approximation can be obtained through the Euler-Lagrange equations applied to the meson fields. This yields a set of coupled, non-linear partial differential equations, expressed in terms of the meson masses, coupling constants, and the nucleon field \citep{MULLER1996508, PhysRevC.63.024314, PhysRevC.65.045201, PhysRevC.68.054318, PhysRevC.70.054309, Kumar2020}. These equations are highly complex and involve various terms and parameters, making them difficult to solve analytically. Therefore, numerical techniques are commonly utilized to derive solutions for the equations and effectively portray the intricate dynamics of mesons in the dense matter system, utilizing the RMF approximation. The expression for conserved nucleon density ($n_{B}$) in this pure baryonic matter can be derived as \citep{MULLER1996508}:
%%%%%%%%%%%%%%%%%%%%%%%%%%%%%%%%%%%%%%%%%%%%%%%%%%%%%%%%%%
\begin{eqnarray}
n_{B} = n_{n} + n_{p} = \frac{k_{F_{n}}^3}{3\pi^2} + \frac{k_{F_{p}}^3}{3\pi^2},
\end{eqnarray}
%%%%%%%%%%%%%%%%%%%%%%%%%%%%%%%%%%%%%%%%%%%%%%%%%%%%%%%%%%
where $k_{F_{n}}^3$ and $k_{F_{p}}^3$ are the fermi momenta of neutrons and protons respectively. To determine the equilibrium state of a neutron star, we must utilize the charge neutrality ($n_{p}=n_{e^{-}}+n_{\mu}$) and beta equilibrium conditions (i.e. in terms of the chemical potential of the particles can be expressed as; $\mu_{n}=\mu_{p}+\mu_{e^{-}}$ \& $\mu_{\mu}=\mu_{e^{-}}$), which offer valuable insight into neutron star behavior and composition. By solving the equations derived from these conditions, we can acquire precise values for energy density and pressure,  which are vital in comprehending its structural properties. Utilizing the charge neutrality and beta equilibrium conditions the relations between the fermi momenta of neutrons, protons, electrons, and muons can be derived as \citep{MULLER1996508, PhysRevC.65.045201}:
%%%%%%%%%%%%%%%%%%%%%%%%%%%%%%%%%%%%%%%%%%%%%%%%%%%%%%%%%%
\begin{eqnarray}
k_{F_{p}} &=& \sqrt[3]{k^3_{F_{e^{-}}} + \sqrt[1.5]{k_{F_{e^{-}}} + M^2_{e^{-}} - M^2_{\mu}}} , \nonumber \\
%k_{F_{\mu}} &=& \sqrt{k_{F_{e^{-}}} + M^2_{e^{-}} - M^2_{\mu}} \nonumber \\
k^2_{F_{n}} &=& k^2_{F_{p}} + k^2_{F_{e^{-}}} + M^2_{e^{-}} + g^2_{\rho} \rho^2 + 4 g_{\sigma} g_{\delta} \sigma \delta - 4 M g_{\delta} \delta  
\nonumber\\
&&
- 2 g_{\rho} \rho \Big(\sqrt{k^2_{F_{p}} + (M - g_{\sigma} \sigma - g_{\delta} \delta)^2} - \sqrt{k^2_{F_{e^{-}}} + M^2_{e^{-}}}\Big)
\nonumber\\
&&
+ 2 \sqrt{[k^2_{F_{p}} + (M - g_{\sigma} \sigma - g_{\delta} \delta)^2] [k^2_{F_{e^{-}}} + M^2_{e^{-}}]}
\end{eqnarray}
%%%%%%%%%%%%%%%%%%%%%%%%%%%%%%%%%%%%%%%%%%%%%%%%%%%%%%%%%%
These relationships account for the internal composition of the star in terms of the chemical potential of the particles. To precisely ascertain the energy density and pressure of the matter within a neutron star using the RMF formalism, it is imperative to consider the interconnections between the Fermi momenta of the constituent particles. However, with the aid of simple calculations involving the energy-momentum tensor and the aforementioned Fermi momentum relations, the numerical values for the energy density and pressure of stellar matter can be readily obtained. The expressions for the energy and pressure can be deduced as follows \citep{Kumar2020}:
%%%%%%%%%%%%%%%%%%%%%%%%%%%%%%%%%%%%%%%%%%%%%%%%%%%%%%%%%%
\begin{eqnarray}\label{edhad}
\epsilon_B & = & \sum_{i=p,n} \frac{g_s}{(2\pi)^{3}}\int_{0}^{k_{F_{i}}} d^{3}k\, \sqrt{k^{2} + (M-g_{\sigma}\sigma- \tau_{3i}g_{\delta}\delta)^2 } 
\nonumber\\
&&
+n_{B} g_\omega\,\omega+m_{\sigma}^2{\sigma}^2\Bigg(\frac{1}{2}+\frac{\kappa_{3}}{3!}\frac{g_\sigma\sigma}{M}+\frac{\kappa_4}{4!}\frac{g_\sigma^2\sigma^2}{M^2}\Bigg)
\nonumber\\
&&
 -\frac{1}{4!}\zeta_{0}\,{g_{\omega}^2}\,\omega^4
 -\frac{1}{2}m_{\omega}^2\,\omega^2\Bigg(1+\eta_{1}\frac{g_\sigma\sigma}{M}+\frac{\eta_{2}}{2}\frac{g_\sigma^2\sigma^2}{M^2}\Bigg)
 \nonumber\\
&&
 + \frac{1}{2} (n_{n} - n_{p}) \,g_\rho\,\rho
 -\frac{1}{2}\Bigg(1+\frac{\eta_{\rho}g_\sigma\sigma}{M}\Bigg)m_{\rho}^2
 \nonumber\\
 && 
-\Lambda_{\omega}\, g_\rho^2\, g_\omega^2\, \rho^2\, \omega^2
+\frac{1}{2}m_{\delta}^2\, \delta^{2}
\nonumber\\
 && 
+\sum_{j=e^{-},\mu}  \frac{g_s}{(2\pi)^{3}}\int_{0}^{k_{F_{j}}} \sqrt{k^2 + m^2_{j}} \, d^{3}k,
\end{eqnarray}
%%%%%%%%%%%%%%%%%%%%%%%%%%%%%%%%%%%%%%%%%%%%%%%%%%%%%%%%
and
\begin{eqnarray}\label{presshad}
p_B & = & \sum_{i=p,n} \frac{g_s}{3 (2\pi)^{3}}\int_{0}^{k_{F_{i}}} d^{3}k\, \frac{k^2}{\sqrt{k^{2} + (M-g_{\sigma}\sigma- \tau_{3i}g_{\delta}\delta)^2}} \nonumber\\
&& - m_{\sigma}^2{\sigma}^2\Bigg(\frac{1}{2} + \frac{\kappa_{3}}{3!}\frac{g_\sigma\sigma}{M} + \frac{\kappa_4}{4!}\frac{g_\sigma^2\sigma^2}{M^2}\Bigg)+ \frac{1}{4!}\zeta_{0}\,{g_{\omega}^2}\,\omega^4 
\nonumber\\
&&
+\frac{1}{2}m_{\omega}^2\omega^2\Bigg(1+\eta_{1}\frac{g_\sigma\sigma}{M}+\frac{\eta_{2}}{2}\frac{g_\sigma^2\sigma^2}{M^2}\Bigg)+\Lambda_{\omega} g_\rho^2 g_\omega^2 \rho^2 \omega^2
\nonumber\\
&&
+ \frac{1}{2}\Bigg(1+\frac{\eta_{\rho}g_\sigma\sigma}{M}\Bigg)m_{\rho}^2\,\rho^{2}-\frac{1}{2}m_{\delta}^2\, \delta^{2}
\nonumber\\
&&
+\sum_{j=e^{-},\mu}  \frac{g_s}{3(2\pi)^{3}}\int_{0}^{k_{F_{j}}} \frac{k^2}{\sqrt{k^2 + m^2_{j}}} \, d^{3}k.
\end{eqnarray}
%%%%%%%%%%%%%%%%%%%%%%%%%%%%%%%%%%%%%%%%%%%%%%%%%%%%%%%%%%
We will now discuss the quarkyonic model along with the relativistic mean field formalism in the context of beta equilibrium and charge neutrality conditions. The theory combined with the quarkyonic model presented in this study suggests that when nucleon densities are low, quarks remain confined within nucleons, and the required interactions for the equation of state of star matter are governed purely by the mean mesonic interacting potential. When the nucleon momenta exceed critical values, referred to as the transition density $n_t$ between hadronic and quarkyonic matter, the low momenta degrees of freedom inside the Fermi sea are treated as non-interacting quarks whereas, at the higher momenta, they are subject to confining forces resulting in the emergence of baryons. The confining interaction becomes dominant near the Fermi surface where momentum exchange is of the order of the QCD confinement scale $\Lambda$. At these momenta, the degrees of freedom near the Fermi surface is confined, and for quark energy scales larger than $\Lambda$, confinement persists. This leads to the formation of a Fermi sea of quarks, which is distinct from the Fermi shell occupied by the nucleons \citep{PhysRevLett.122.122701}. The quarks that exist within the Fermi sea are considered to be effectively non-interacting, primarily because of the Pauli exclusion principle. On the other hand, the nucleons experience strong interactions, and these interactions can be accurately described by the relativistic mean-field formalism. 

In the quarkyonic matter proposed by McLerran and Reddy, and later extended by Zhao and Lattimer for beta equilibrated matter \citep{PhysRevD.102.023021}, the two flavors of quarks, d and u, are taken into account. These quarks form a Fermi sea, while the nucleons form a Fermi shell occupying the minimum momentum states $k_{0_{(n,p)}}$ and Fermi momentum $k_{F_{(n,p)}}$. The low momentum degrees of freedom inside the Fermi sea behave as non-interacting quarks, filling up the momentum states with $k_{F_{u}}$ and $k_{F_{d}}$, being the Fermi momentum of up and down quarks respectively. The total baryon density and charge neutrality for such a system take into account the contributions from both the nucleons and quarks and ensure that the total baryon density is conserved while maintaining charge neutrality, which can be expressed as,
%%%%%%%%%%%%%%%%%%%%%%%%%%%%%%%%%%%%%%%%%%%%%%%%%%%%%%%%%%
\begin{eqnarray}
n &=& n_n + n_p + \frac{n_u+n_d}{3}
\nonumber\\
&=& \frac{g_s}{6\pi^2}\bigg[(k_{F_{n}}^3-k_{0_{n}}^3)+(k_{F_{p}}^3-k_{0_{p}}^3)+\frac{(k_{F_{u}}^3+k_{F_{d}}^3)}{3}\bigg],  
\end{eqnarray}
%%%%%%%%%%%%%%%%%%%%%%%%%%%%%%%%%%%%%%%%%%%%%%%%%%%%%%%%%%
and
%%%%%%%%%%%%%%%%%%%%%%%%%%%%%%%%%%%%%%%%%%%%%%%%%%%%%%%%%%
\begin{eqnarray}
n_p + \frac{2n_{u}-n_{d}}{3} & = & n_{e^{-}} + n_{\mu},  
\end{eqnarray}
%%%%%%%%%%%%%%%%%%%%%%%%%%%%%%%%%%%%%%%%%%%%%%%%%%%%%%%%%%
where $n_u$ and $n_d$ being the number density of up and down quarks and $g_s=2$ is the fermionic spin degeneracy. In the work by Zhao and Lattimer, they established a relation between the minimum allowed momentum near the Fermi surface ($k_{0_{(n,p)}}$), the transition Fermi momentum ($k_{t_{(n,p)}}$), which represents the momentum at which quarks start to emerge, and the Fermi momentum ($k_{F_{(n,p)}}$) of neutrons and protons in $\beta$-equilibrated (neutron star) matter. This relationship can be written as,
%%%%%%%%%%%%%%%%%%%%%%%%%%%%%%%%%%%%%%%%%%%%%%%%%%%%%%%%%%
\begin{eqnarray}
 k_{0(n,p)} &=& (k_{F_{(n,p)}}-k_{t_{(n,p)}})\bigg[1+ \frac{\Lambda^2}{k_{F_{(n,p)}}k_{t_{(n,p)}}}\bigg]
\end{eqnarray}
%%%%%%%%%%%%%%%%%%%%%%%%%%%%%%%%%%%%%%%%%%%%%%%%%%%%%%%%%%
has significant implications of the chemical equilibrium in neutron stars, particularly in the context of the coexistence of quarks and nucleons. This equation offers valuable insights into the intricate interplay between quarks and nucleons under the extreme conditions that exist within neutron stars and is critical for developing accurate models of the behavior of matter in neutron stars. In order to constrain the value of $\Lambda$, Zhao and Lattimer utilized empirical values of nuclear saturation density and slope parameter for symmetric and pure nuclear matter. They determined that the optimal values for Lambda are $800$ and $1400$ $MeV$ at different transition densities. In accordance with their findings, we have also adopted the same values of Lambda in all our simulations so that we can ensure our results are consistent with the saturation properties of nuclear matter.  Now, for the theoretical  exploration of astrophysical quantities, the principle of strong interaction equilibrium plays a pivotal role in determining the optimal distribution of particle concentrations. This principle necessitates the minimization of the total energy in relation to the particle concentrations while maintaining a constant density and lepton fraction. The principle of strong interaction equilibrium can be viewed as analogous to the concept of chemical potential equilibrium between nucleons and quarks, resulting in,
%%%%%%%%%%%%%%%%%%%%%%%%%%%%%%%%%%%%%%%%%%%%%%%%%%%%%%%%%%
\begin{eqnarray}\label{ebnq}
\mu_n &=& \mu_u+2\mu_d,\nonumber\\
\mu_p &=& 2\mu_u+\mu_d,
\end{eqnarray}
%%%%%%%%%%%%%%%%%%%%%%%%%%%%%%%%%%%%%%%%%%%%%%%%%%%%%%%%%%
with $\mu_u$ and $\mu_d$ being the chemical potentials of the up and down flavor quarks respectively. Furthermore, the principle of beta equilibrium demands the additional minimization of the total energy density in relation to the lepton concentrations while maintaining a constant baryon density. It is applicable when the timescales of weak interactions are shorter than the dynamical timescales. This principle is essential and in addition to our previously defined relations between the chemical potential of nucleons and leptons, 
%%%%%%%%%%%%%%%%%%%%%%%%%%%%%%%%%%%%%%%%%%%%%%%%%%%%%%%%%%
\begin{eqnarray} \label{qnbe}
\mu_{n} &=& \mu_{p}+\mu_{e^{-}} , \nonumber \\
\mu_{\mu} &=& \mu_{e^{-}},
\end{eqnarray}
%%%%%%%%%%%%%%%%%%%%%%%%%%%%%%%%%%%%%%%%%%%%%%%%%%%%%%%%%%
it establishes an additional relation between quarks and leptons for quarkyonic matter, i.e,
%%%%%%%%%%%%%%%%%%%%%%%%%%%%%%%%%%%%%%%%%%%%%%%%%%%%%%%%%%
\begin{eqnarray} \label{qlbe}
\mu_{e^-} = \mu_{\mu} = \mu_d-\mu_u.   
\end{eqnarray}
%%%%%%%%%%%%%%%%%%%%%%%%%%%%%%%%%%%%%%%%%%%%%%%%%%%%%%%%%%
Another important aspect of this quarkyonic model is the demand that both quark flavors exist at the transition density, referred to as $n_t$, implying that the quark masses are not independent variables in the model. Instead, their values are dictated by the prevailing beta-equilibrium conditions at $n_t$, which are reliant on the nucleon potential \citep{PhysRevD.102.023021}. The present study relies on the RMF formalism to furnish the nucleon potential at the point of transition density, as illustrated by equations \ref{edhad} and \ref{presshad}. By utilizing equation \ref{ebnq} at the transition density, we can ascertain the mass of up and down quarks, which will lead us to the following expressions,  
%%%%%%%%%%%%%%%%%%%%%%%%%%%%%%%%%%%%%%%%%%%%%%%%%%%%%%%%%%
\begin{eqnarray}
 m_{u} = \frac{2}{3} \mu_{t_{p}} -  \frac{1}{3} \mu_{t_{n}}, \,\,\,\,\,\,\,\,\,\,\,\,   m_{d} = \frac{2}{3} \mu_{t_{n}} -  \frac{1}{3} \mu_{t_{p}},
\end{eqnarray}
%%%%%%%%%%%%%%%%%%%%%%%%%%%%%%%%%%%%%%%%%%%%%%%%%%%%%%%%%%
 where $\mu_{t_{n}}$ and $\mu_{t_{p}}$ will be given by,
%%%%%%%%%%%%%%%%%%%%%%%%%%%%%%%%%%%%%%%%%%%%%%%%%%%%%%%%%%
\begin{eqnarray}
\mu_{t_{n}} &=& \sqrt{k_{t_{n}}^{2} + (M - g_{\sigma}\sigma - g_{\delta}\delta)^2} - g_{\omega}\omega - \frac{g_{\rho}}{2}\rho, \nonumber \\
\mu_{t_{p}} &=& \sqrt{k_{t_{p}}^{2} + (M - g_{\sigma}\sigma + g_{\delta}\delta)^2} - g_{\omega}\omega + \frac{g_{\rho}}{2}\rho
\end{eqnarray}
%%%%%%%%%%%%%%%%%%%%%%%%%%%%%%%%%%%%%%%%%%%%%%%%%%%%%%%%%%
 and the values of $\sigma$, $\omega$, $\rho$, and $\delta$ should also be obtained at the corresponding transition densities of neutron and proton respectively. Having obtained all the relevant expressions and relationships, we are now able to compute the Fermi momentum of up and down quarks for use in numerical simulations, which can be derived using equations \ref{qnbe} and \ref{qlbe}.  
%%%%%%%%%%%%%%%%%%%%%%%%%%%%%%%%%%%%%%%%%%%%%%%%%%%%%%%%%%
\begin{figure}
\centering
\includegraphics[width=1\columnwidth]{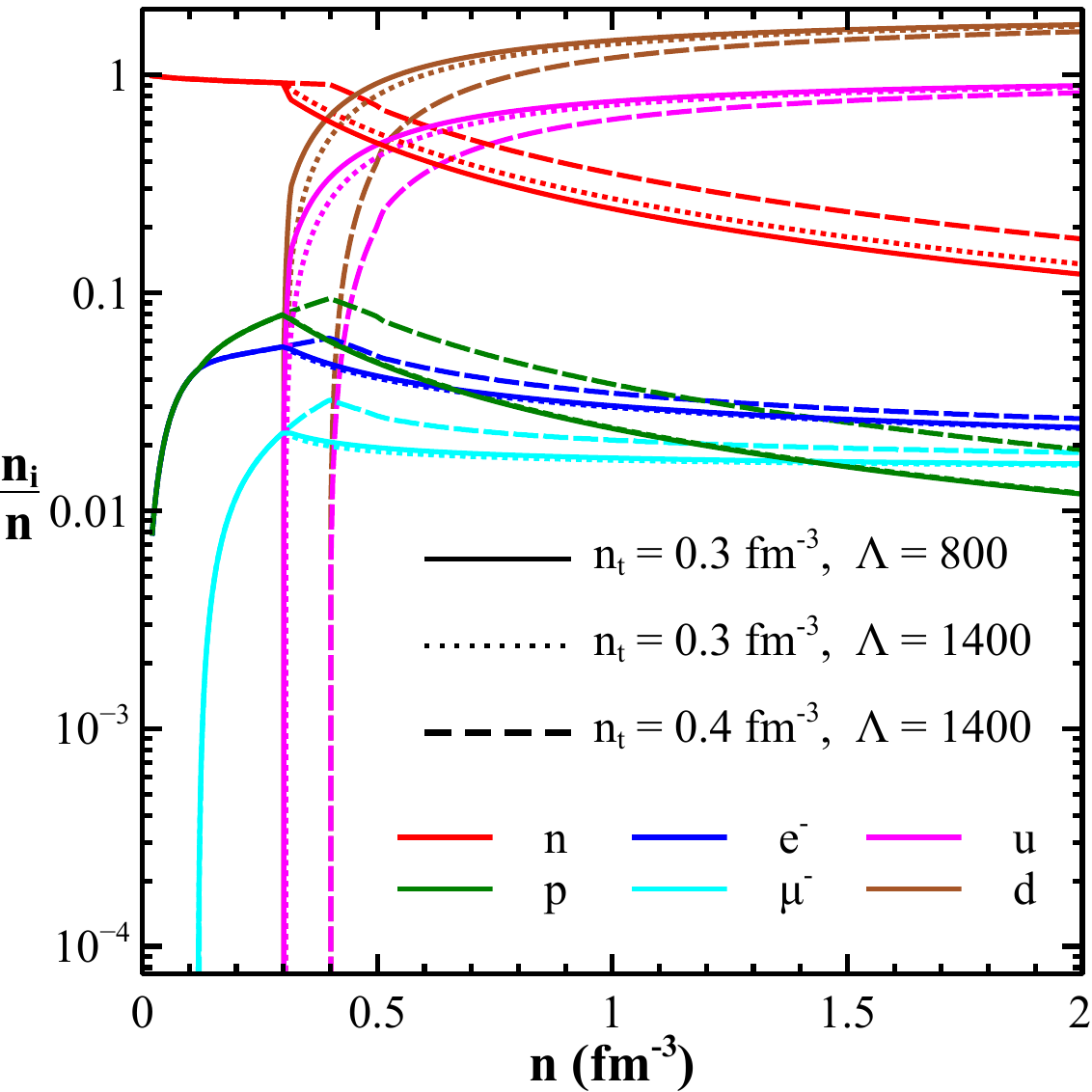}
\caption{The particle fractions for different combinations of transition densities and $\Lambda$ using the G3 as RMF parameter set for beta-equilibrated quarkyonic neutron star matter. The solid lines correspond to ($n_t = 0.3\, fm^{-3}$, $\Lambda = 800\, MeV$), the dashed lines represent ($n_t = 0.3\, fm^{-3}$, $\Lambda = 1400\, MeV$), and the dotted lines depict the ($n_t = 0.4\, fm^{-3}$, $\Lambda = 1400\, MeV$) case.}
\label{fig1}
\end{figure}
%%%%%%%%%%%%%%%%%%%%%%%%%%%%%%%%%%%%%%%%%%%%%%%%%%%%%%%%%%
After considering all the expressions for beta and chemical equilibrium, we calculated the particle fractions of nucleons, leptons, and quarks in neutron star matter, which are presented in Figure \ref{fig1}. For density greater than the transition value, the presence of quarks leads to a significant transformation in the composition of star matter. We have examined two distinct scenarios involving transition densities of $0.3\, fm^{-3}$ and $0.4\, fm^{-3}$. For $n_{t} = 0.3\, fm^{-3}$, we used $\Lambda$ values of $800$ and $1400\, MeV$, and for the $n_{t} = 0.4\, fm^{-3}$ case, we only used the $\Lambda$ value of $1400\, MeV$, as previously mentioned. For the baryonic interactions, we use G3 as the RMF parameter set to supply the required values of coupling constants within the frozen-in mean mesonic field. After the transition density, when quarks come into existence, there is a noticeable decrease in the fraction of neutrons and protons found within the matter for all the adopted scenarios. Upon reaching the transition density $n_t$, there is a significant rise in both up and down quark fractions to maintain the charge neutrality in the system, with a noticeable dominance of the down quark population over the up quark population. The resulting state of matter fundamentally alters the properties of matter such as its equation of state, is highly compressed and dense, and can exhibit exotic phenomena, such as color superconductivity and the formation of strange matter.

\begin{figure}
\centering
\includegraphics[width=1\columnwidth]{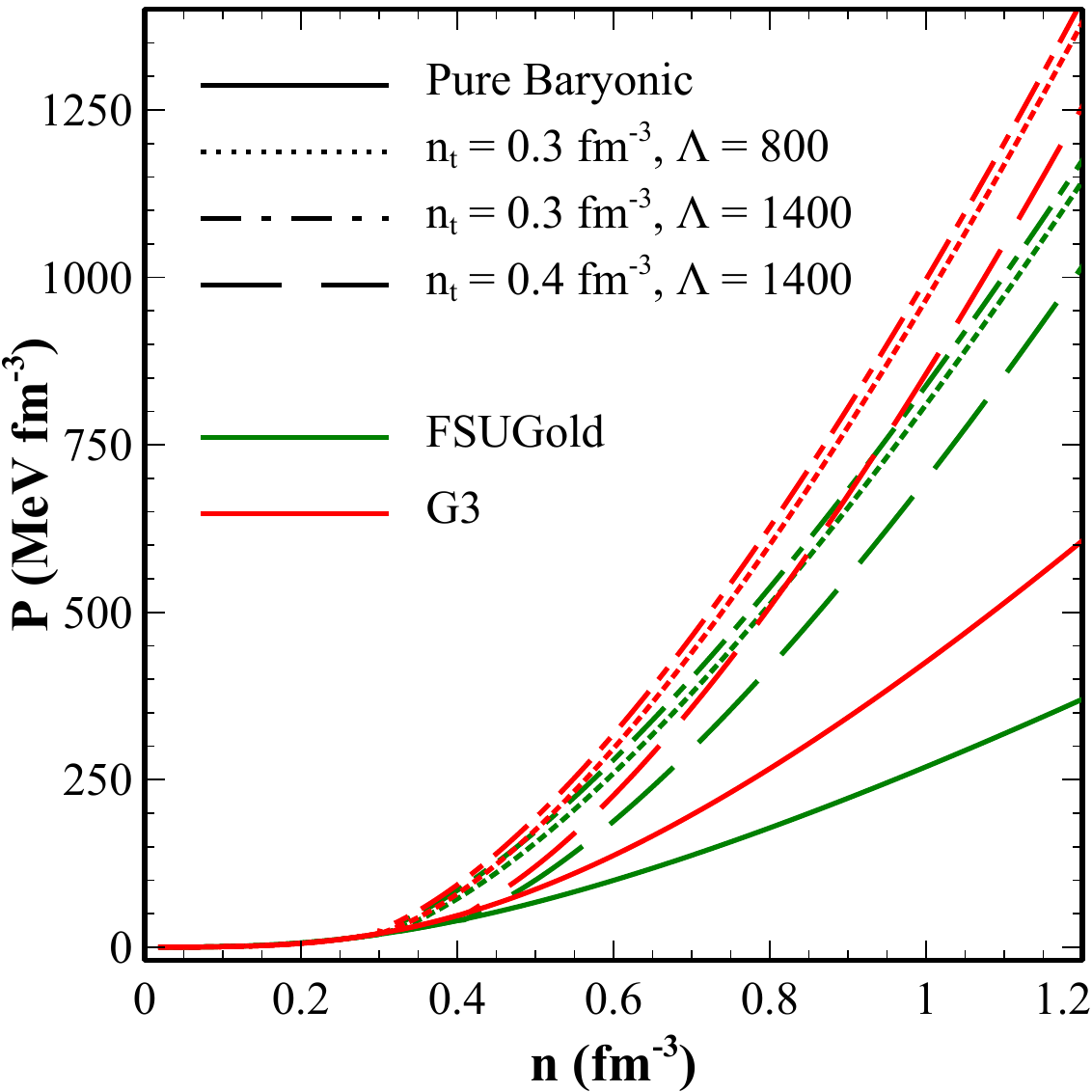}
\caption{Equation of state (Pressure as a function of number density) for pure baryonic and quarkyonic stellar matter using FSUGold and G3 RMF parameter sets. The Green color lines correspond to the FSUGold and the red color lines represent the G3 parameter set equation of states. The solid line for both the parameter sets represents the pressure as a function of number density for the pure baryonic stellar matter (without quarks). The dotted, dash-dot-dash, and dashed lines represent the EOS of star matter with the quarkyonic model for $(n_{t} = 0.3\, fm^{-3}, \Lambda = 800\, MeV)$, $(n_{t} = 0.3\, fm^{-3}, \Lambda = 1400\, MeV)$, and $(n_{t} = 0.4\, fm^{-3}, \Lambda = 1400\, MeV)$ respectively.}
\label{fig2}
\end{figure}

Now, the total energy density of the quarkyonic star matter with interacting nucleons via mesons, leptons and the non-interacting quarks will be given by,
%%%%%%%%%%%%%%%%%%%%%%%%%%%%%%%%%%%%%%%%%%%%%%%%%%%%%%%%%%
\begin{eqnarray}
\varepsilon &=& \epsilon_B + \epsilon_u + \epsilon_d,
\end{eqnarray}
%%%%%%%%%%%%%%%%%%%%%%%%%%%%%%%%%%%%%%%%%%%%%%%%%%%%%%%%%%
where $\epsilon_B$ denotes the energy density for interacting nucleons and also for non-interacting leptons. The determination of $\epsilon_B$ in the study of hybrid quarkyonic matter depends on the transition density, a point at which the system undergoes a change in behavior. Before reaching the transition density, the value of $\epsilon_B$ can be obtained simply by applying equation \ref{edhad}. However, once the transition density is reached, this expression must be modified. In particular, $k_{0_{n,p}}$, the lowest possible momentum of neutrons/protons, replaces zero as the integration limit in the first term of equation \ref{edhad} (i.e. the integration will be $\int_{k_{0_{i}}}^{k_{F_{i}}}$). The mean mesonic fields ($\sigma, \omega, \rho, \delta$) are also adjusted simultaneously to match the interaction potential with the permissible Fermi momentum of nucleons. All of these factors are carefully considered and accounted for in our numerical simulations of quarkyonic star matter. Apart from this, the crust of a neutron star is also an important and complex region, and its properties can have a significant impact on the overall equation of state. In order to ensure that our equation of state is as comprehensive as possible, we have added the crustal data to the baryonic energy density part. This data was borrowed from our previous work on the crustal properties of the neutron star with RMF formalism \citep{PhysRevD.105.043017}. By incorporating the data we have gathered on the crust, we can better understand the behavior of the neutron star as a whole and generate a more complete and accurate equation of state. However, the energy density for the non-interacting quarks can be obtained by,
%%%%%%%%%%%%%%%%%%%%%%%%%%%%%%%%%%%%%%%%%%%%%%%%%%%%%%%%%%
\begin{eqnarray}
\epsilon_{u} + \epsilon_{d} &=& \sum_{j=u,d}\frac{g_s N_c}{(2\pi)^3}\int_0^{k_{F_{j}}}k^2\sqrt{k^2 + m_{j}^2 }\, d^3k.
\end{eqnarray}
%%%%%%%%%%%%%%%%%%%%%%%%%%%%%%%%%%%%%%%%%%%%%%%%%%%%%%%%%%
where $g_{s}$ is the spin degeneracy factor and $N_{c}$ is the quark color degeneracy. In contrast to the assumption made by Zhao and Lattimer in their study, where they assumed that the potential energy of nucleons remains constant above and below the transition density $n_{t}$, in this study, the potential energy is determined by the actual microscopic potential which is dependent on the density of nucleons and the compositional structure of the stellar matter, as evident from equation \ref{edhad}. After determining the energy density, we can proceed to obtain the pressure of the quarkyonic neutron star matter. However, for pure baryonic matter, the pressure can be obtained using equation \ref{presshad}, and for the quarkyonic matter, it is calculated using,
%%%%%%%%%%%%%%%%%%%%%%%%%%%%%%%%%%%%%%%%%%%%%%%%%%%%%%%%%%
\begin{eqnarray}
p &=& p_{B} + \mu_{u} n_{u} + \mu_{d} n_{d} - \epsilon_{u} - \epsilon_{d} 
\end{eqnarray}
%%%%%%%%%%%%%%%%%%%%%%%%%%%%%%%%%%%%%%%%%%%%%%%%%%%%%%%%%%
In this context, the symbol $p_B$ represents the pressure of the hybrid matter arising from the baryons and leptons after the transition density has been reached, which can be expressed by means of equation \ref{presshad}, taking into account the impact of the minimum permissible momentum of the nucleons present in the quark fermi sea (similarly like the case for $\epsilon_{B}$ expression after $n_{t}$ is achieved). To facilitate deeper investigations into the dynamics of static and rotating neutron stars, this study involves computing the equation of state, encompassing number density, energy density, and pressure, for quarkyonic star matter in three distinct scenarios, as well as for the pure baryonic star matter using the FSUGold \citep{PhysRevLett.95.122501} and G3  \citep{KUMAR2017197} parameter sets of the RMF formalism. The variation of pressure as a function of number density for the pure baryonic and (hybrid) quarkyonic matter with three different scenarios i.e. EOS-q1 $(n_{t} = 0.3\, fm^{-3}, \Lambda = 800\, MeV)$, EOS-q2 $(n_{t} = 0.3\, fm^{-3}, \Lambda = 1400\, MeV)$, and EOS-q3 $(n_{t} = 0.4\, fm^{-3}, \Lambda = 1400\, MeV)$, with both the parameter sets is depicted in Fig. \ref{fig2}; these three cases will be referred to as EOS-q1, EOS-q2, and EOS-q3 in subsequent discussions. 
%%%%%%%%%%%%%%%%%%%%%%%%%%%%%%%%%%%%%%%%%%%%%%%%%%%%%%%%%%
%%%%%%%%%%%%%%%%%%%%%%%%%%%%%%%%%%%%%%%%%%%%%%%%%%%%%%%%%%
We observed that the quarkyonic model leads to the stiffer EOS in all scenarios as compared to the pure baryonic case. The pressure curve for pure baryonic matter, composed solely of nucleons, exhibits a smooth increase in pressure with the density. However, the introduction of quark matter causes the pressure to become discontinuous at the transition density of $0.3\, fm^{-3}$ and $0.4\, fm^{-3}$, leading to a stiffer equation of state in comparison to the pure baryonic case, which holds true for both of the proposed sets of RMF parameters. This crossover transition between nucleons and quarks differs from the conventional first-order phase transition studied so far, where the pressure decreases and results in a softer equation of state \citep{PhysRevD.46.1274, PhysRevC.81.045201,refId0, Prasad2018-ck}. This abrupt jump in pressure is a distinguishing feature of the quarkyonic model, caused by the presence of deconfined quarks occupying lower momentum states within the core of the star, which leads to a substantial increase in the pressure of the system. As a consequence, for a specific density, we encounter elevated pressure that can sustain a comparatively more massive neutron star, in contrast to the scenario of pure baryonic degrees of freedom. The stiff equation of states that are acquired through the quarkyonic model has many significant ramifications for the examination of neutron stars. This is because it has the potential to influence their configuration and evolution, which we will explore in the following sections of this paper. In addition, while conducting our calculations, we noticed that for the quarkyonic matter EOS at high density, the pressure tends to approach the asymptotic limit value of $\epsilon/3$, accounting for the validation of the model.
%%%%%%%%%%%%%%%%%%%%%%%%%%%%%%%%%%%%%%%%%%%%%%%%%%%%%%%%%%
%%%%%%%%%%%%%%%%%%%%%%%%%%%%%%%%%%%%%%%%%%%%%%%%%%%%%%%%%%
%%%%%%%%%%%%%%%%%%%%%%%%%%%%%%%%%%%%%%%%%%%%%%%%%%%%%%%%%%
\section{Static NS}\label{Static}
%%%%%%%%%%%%%%%%%%%%%%%%%%%%%%%%%%%%%%%%%%%%%%%%%%%%%%%%%%
%%%%%%%%%%%%%%%%%%%%%%%%%%%%%%%%%%%%%%%%%%%%%%%%%%%%%%%%%%
%%%%%%%%%%%%%%%%%%%%%%%%%%%%%%%%%%%%%%%%%%%%%%%%%%%%%%%%%%
In this particular section, we will utilize the equation of states that was developed in the earlier section to delve into the properties of a static neutron star. The scrutiny of observational data on inspirals of compact stars from LIGO/Virgo offers a significant means to authenticate the EOS of dense matter at the densities that compact stars occupy. The investigation and the characterization of hybrid quark star EOSs through gravitational-wave data is a relatively nascent field and can provide insight into the properties of these exotic stars, including their size, mass, and stiffness of their EOS. The constraints inferred from observational data, coupled with causality deliberations, strongly suggest that the speed of sound calculated using the EOS of highly dense quark matter regime found in neutron stars ought to approximate the relativistic value of $c/\sqrt{3}$, $c$ being the speed of light. The complex internal structure of a neutron star results in non-uniformity in its composition, leading to different trends in the speed of sound across its crust and core regions. 
%%%%%%%%%%%%%%%%%%%%%%%%%%%%%%%%%%%%%%%%%%%%%%%%%%%%%%%%%%
\begin{figure}
\centering
\includegraphics[width=1\columnwidth]{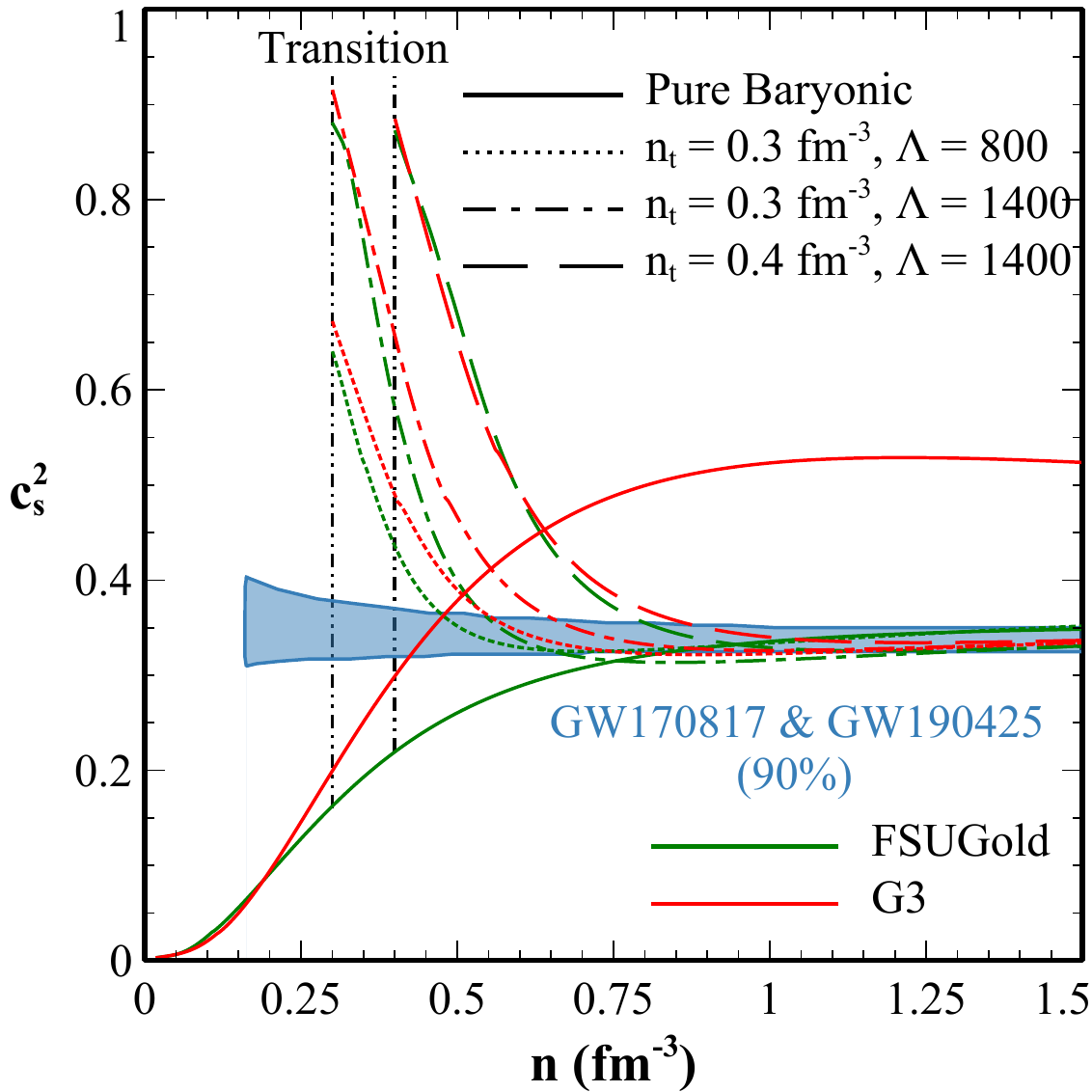}
\caption{Speed of sound ($c_s^2$) as a function of density for the pure baryonic and different groups of quarkyonic matter equation of states. The Green color lines represent the $c_s^2$ values for FSUGold, and the red color portrays the G3 parameter set. The dotted, dash-dot-dash, and dashed lines represent three different scenarios for a neutron star with quarkyonic matter (same as Fig. \ref{fig2}). The shaded area in blue illustrates the posterior distributions of $c_s^2$, as inferred from the joint observational analysis of GW170817 and GW190425 data, with a $90\,\%$ confidence level.}
\label{fig3}
\end{figure}
%%%%%%%%%%%%%%%%%%%%%%%%%%%%%%%%%%%%%%%%%%%%%%%%%%%%%%%%%%
Within the crust, the velocity of sound is influenced by the composition, shapes, and arrangement of atomic nuclei, which can significantly impact the star's overall crustal rigidity. This crucial factor is incorporated in this study by including the crustal EOS from \cite{PhysRevD.105.043017}. It is apparent from Fig. \ref{fig2} that the G3 parameter set for pure baryonic stars yields a stiffer EOS compared to the FSUGold parameter set. This implies that the G3 EOS predicts a higher mass for a pure baryonic neutron star, as illustrated in Fig. \ref{fig4}. However, while pure baryonic stars with the G3 parameter set are able to achieve a mass close to $2\,M_\odot$ (solar mass), they fail to satisfy the QCD asymptotic freedom causality limit of the speed of sound in the highly-dense astrophysical matter \citep{PhysRevD.95.044032}. On the other hand, the pure baryonic star with FSUGold EOS remains consistent with the causality limit of sound speed in compact stars, unlike the G3 parameter set. However, it should be noted that the mass predicted by the FSUGold EOS for a neutron star is comparatively low. While the consistency of the FSUGold pure baryonic EOS with the causality limit is a significant advantage, the relatively low mass predictions may limit its applicability in certain astrophysical contexts. Obtaining a neutron star with a mass greater than $2\, M_{\odot}$ while satisfying the causality criteria and other observational constraints has been a persistent challenge for the parametrizations of RMF formalism \citep{PhysRevC.98.065804, 10.1093/mnras/stz654}. The speed of sound computed in this study using $c_s^2 = \frac{\partial p}{\partial \epsilon}$ from the EOSs derived for pure baryonic and quarkyonic star matter using both the RMF parameter sets under examination is illustrated in Fig. \ref{fig3}. The posterior distributions of $c_s^2$, as determined by the Bayesian Inference from the joint observational analysis of GW170817 \citep{2017PhRvL.119p1101A} and GW190425 \citep{2020ApJ...892L...3A} data, have been illustrated using the shaded area in Fig. \ref{fig3}. This analysis was conducted by Miao et. al, with a high degree of certainty ($90\,\%$ confidence level) by considering the strange quark stars as the merging components \citep{Miao_2021}. Fig. \ref{fig3} also provides insight into the behavior of the speed of sound as determined by various quarkyonic star matter equations of states. We observe a significant spike in the speed of sound at the transition density due to inconsistencies in the uniformity of matter caused by the emergence of quarks. However, despite these fluctuations, we find that all quarkyonic star matter EOSs (i.e. EOS-q1, EOS-q2, EOS-q3), for both the RMF parameter sets, align with the constraints established by the analysis of observational gravitational wave data at high densities. The emergence of quarks in the transition density region of quarkyonic stars, due to the unique properties of the ultra-dense matter present in these areas, can lead to phenomena that are not well understood. Various studies investigating statistical analyses pertaining to the speed of sound in neutron stars have also demonstrated a comparable pattern, exhibiting a peak around 2-3 times the nuclear saturation density \citep{Tews_2018, Altiparmak_2022}. These studies have employed different methodologies, including Bayesian inference, Monte Carlo simulations, and Maximum Likelihood estimation, to analyze observational data and extract insights into the physical properties of ultra-dense matter. Nonetheless, the agreement between the obtained EOSs and gravitational wave data (GW170817 and GW190425) provides significant support for the potential existence of quarkyonic stars, as well as the validity of the model used to describe them.
%%%%%%%%%%%%%%%%%%%%%%%%%%%%%%%%%%%%%%%%%%%%%%%%%%%%%%%%%%
\begin{figure}
\centering
\includegraphics[width=1\columnwidth]{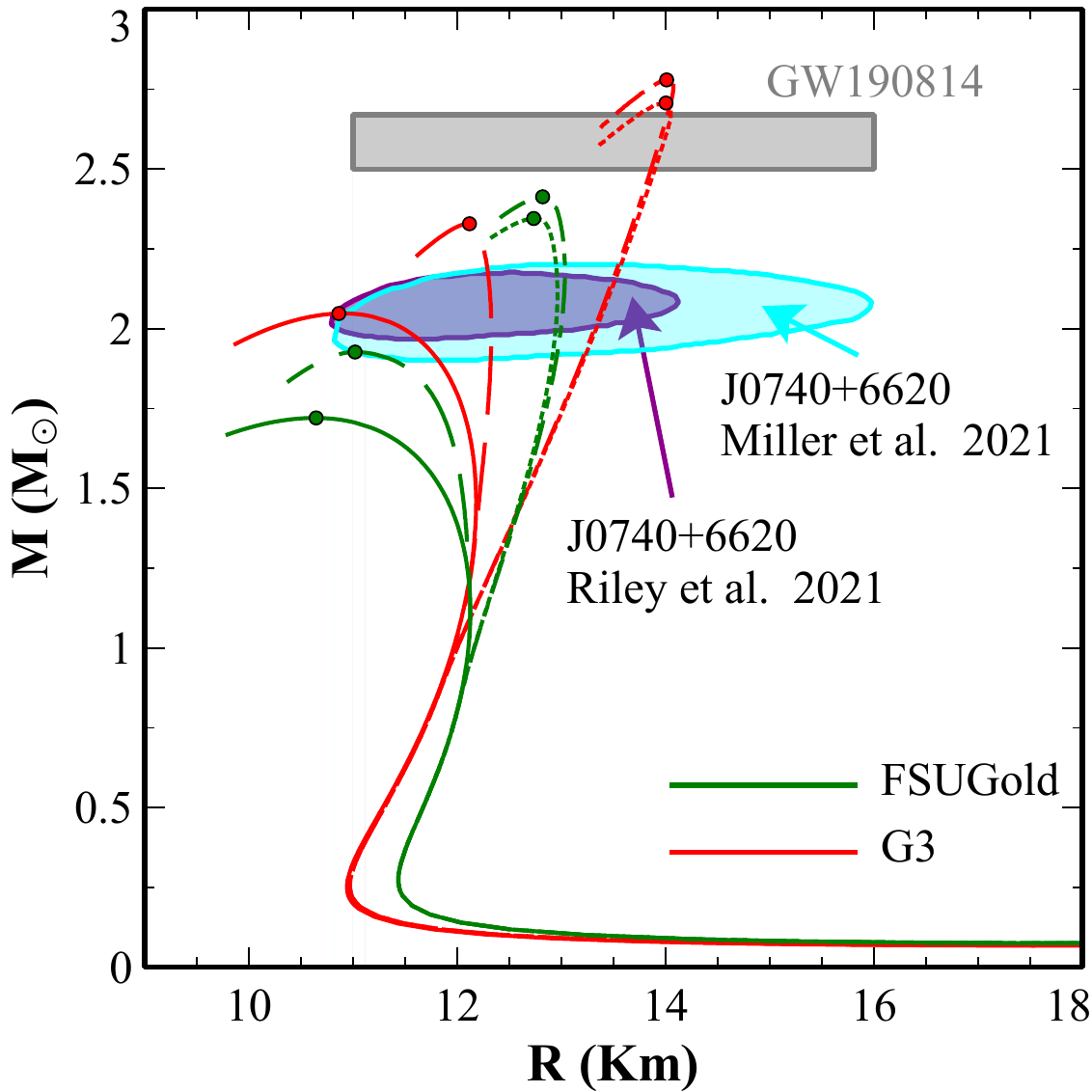}
\caption{Mass-Radius (M-R) curve for pure baryonic and quarkyonic star matter. The solid lines correspond to the M-R profile of a neutron star composed entirely of baryonic matter; the dotted, dash-dot-dashed, and dashed lines represent the Mass-Radius curves for EOS-q1, EOS-q2, and EOS-q3 of a neutron star consisting of beta-equilibrated quarkyonic matter (i.e. values of $n_t$ and $\Lambda$ are same as for Fig. \ref{fig2}). The shaded regions represent the constraints determined by analyzing observational data from the corresponding gravitational events.}
\label{fig4}
\end{figure}
%%%%%%%%%%%%%%%%%%%%%%%%%%%%%%%%%%%%%%%%%%%%%%%%%%%%%%%%%%

Now the mass ($M$) and radius ($R$) of a static and isotropic neutron star for the obtained EOSs are determined by solving a set of coupled differential equations, which are collectively known as the TOV equations \citep{PhysRev.55.364, PhysRev.55.374}, 
%%%%%%%%%%%%%%%%%%%%%%%%%%%%%%%%%%%%%%%%%%%%%%%%%%%%%%%%%%
\begin{eqnarray}
\frac{dp(r)}{dr} &=& - \frac{[p(r)+{\epsilon(r)}][m(r)+4\pi r^3 p(r)]}{r[r-2m(r)]}, \\
\frac{dm(r)}{dr} &=& 4\pi r^2 {\epsilon}(r),
\end{eqnarray}
%%%%%%%%%%%%%%%%%%%%%%%%%%%%%%%%%%%%%%%%%%%%%%%%%%%%%%%%%%
In this study, the mass-radius relationship of quarkyonic stars has been examined through the solution of the TOV equations using appropriate initial and boundary conditions. The results obtained demonstrate a striking contrast between quarkyonic and pure baryonic stars, as evidenced by the mass-radius profile depicted in Fig. \ref{fig4}. The EOS for quarkyonic star matter obtained utilizing both the RMF parameter sets and exploring various scenarios with a range of values for transition density ($n_t$) and the parameter ($\Lambda$), predicts exceptionally high maximum mass and significantly larger corresponding radii in comparison to pure baryonic stars. The mass-radius profile for quarkyonic star matter itself exhibits a remarkable sensitivity to changes in the values of both of its free parameters, namely the transition density, and the parameter $\Lambda$. Also, given the sensitivity of the equation of state to the variation of both the transition density and the parameter $\Lambda$, it is essential to understand how changes in these parameters affect the properties of quarkyonic stars.  We know that the transition density is a fundamental parameter that determines the onset of the phase transition between hadronic and quark matter. Hence, the value of $n_t$ plays a critical role in determining the properties of quarkyonic stars. As the density of the star increases beyond the transition density, the presence of quarks causes the equation of state to become increasingly stiffer (illustrated in Fig. \ref{fig2}), leading to an increase in the maximum mass and radius of the star. The results obtained from our calculations clearly demonstrate a significant difference in the maximum mass and corresponding radii of the star when using EOS-q1 and EOS-q2, which share the same values for transition density, in comparison to EOS-q3 ($n_t = 0.4\, fm^{-3}$). This trend is observed for both G3 and FSUGold parameter sets. This emphasizes the critical nature of precisely determining the transition density in quarkyonic star models. Given that $n_{t}$ is regarded as a free parameter in the current model, it is possible to significantly constrain the density limit for the appearance of quarks by leveraging gravitational observational data. Similarly, the parameter $\Lambda$ is another essential free parameter that appears in the quarkyonic star model. In the broad sense, it is related to the scale of chiral symmetry breaking in the effective field theory used to describe the quarkyonic matter. The value of Λ plays a pivotal role in influencing the behavior of quarkyonic matter at high densities, resulting in notable fluctuations in the equation of state. A rise in the value of $\Lambda$, while maintaining a constant value for $n_{t}$, leads to an increase in the maximum mass exhibited by the star. This variation can be accurately demonstrated by examining the mass-radius profile of quarkyonic star matter, obtained from EOS-q1 and EOS-q2 of the corresponding RMF parameter, which differs solely in terms of the value of $\Lambda$.

\begin{figure}
\centering
\includegraphics[width=1\columnwidth]{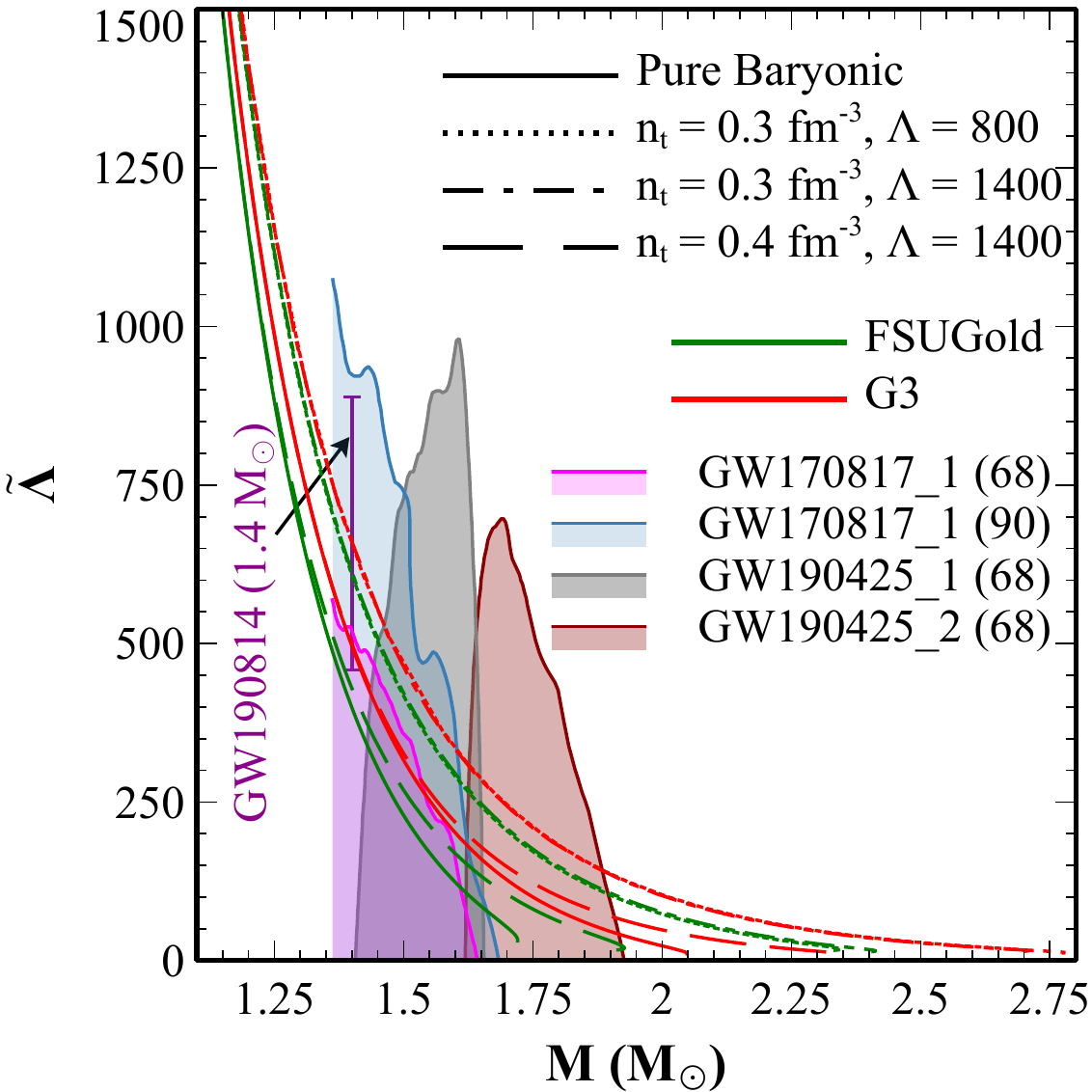}
\caption{This plot displays the relationship between the tidal deformability $\tilde{\Lambda}$ and the maximum mass of the star. The shaded region exhibit posterior distribution of tidal deformability from various gravitational wave events with their respective credible level while the estimated range of tidal deformability for a canonical star obtained from the analysis of observational GW190814 data is shown with hatched line. The blue and magenta shaded regions represent the constraints on the heavier component of GW170817 at $68\%$ and $90\%$ CI respectively. The grey and brown shaded regions depict the limits for the binary components of GW190425 event at a $68\%$ CI.}
\label{fig5}
\end{figure}

The restrictions imposed on the upper bounds of both the mass and radius of neutron stars, as inferred from the analysis of numerous observational phenomena, are also illustrated in Fig. \ref{fig4}. It has been noted that the pure baryonic FSUGold EOS yields a substantially lower mass for the neutron star. Only the maximum mass and radius projected by the G3 pure baryonic EOS aligns with the observational constraints presented by Riley et. al. and Miller et. al. for the PSR J0740+6620 pulsar \citep{2021ApJ...915L..12F, Riley_2021, 2021ApJ...918L..28M}. As previously stated, there exists a contradiction since FSUGold pure baryonic EOS can satisfy the conformal limit for the velocity of sound at high density but is unable to forecast the higher mass of neutron stars. Conversely, the G3 pure baryonic EOS can predict higher masses but falls short of meeting the conformal limit for the velocity of sound at high density. We can potentially resolve this situation by incorporating quarks, as evidenced by Fig. \ref{fig4}. The quarkyonic star EOS-q3, which is based on the FSUGold parameter set, predicts mass range and sound speed that falls within the observational limits of the PSR J0740+6620 pulsar. Recent observations of the "black widow" pulsar, PSR J0952-0607, which is currently recognized as the fastest and most massive neutron star known, have indicated that its estimated mass is $M \sim 2.35 \pm 0.17 M_{\odot}$ \citep{Romani_2022}. These findings suggest that it could potentially be considered a candidate for quarkyonic star classification. Our findings reveal that the maximum mass calculated for the quarkyonic star using the EOS-q1 and EOS-q2 from the FSUGold parameter set, as well as the EOS-q3 from the G3 parameter set, falls perfectly within the range of the black widow pulsar, providing additional evidence for the existence of quarkyonic stars. In the event that forthcoming observations reveal compact stars with masses surpassing $2.6 M_{\odot}$, current theoretical models, such as pure hadronic RMF, spectral decomposition, piecewise polytropic, etc., would be unable to predict such high masses while simultaneously surpassing the required QCD conformal limit of the speed of sound within a dense system. The GW190814 gravitational observation event, which resulted from the merger of a black hole (primary component) and a secondary component with a mass of approximately 2.6$M_\odot$, is notable for providing evidence of the existence of massive compact objects \citep{Abbott2020}. The absence of an electromagnetic counterpart to the GW190814 gravitational observational event presents a significant challenge in determining the true nature of its secondary component. This component may potentially be the lightest black hole or the heaviest neutron star, leading to considerable difficulty in explaining the characteristics of this compact object. As a result, there is much debate surrounding the mystery of the secondary component in the GW190814 event, and a comprehensive understanding of its nature remains elusive. Specifically, our theoretical analysis indicates that the secondary component in the GW190814 event could potentially be a quarkyonic star. This conclusion is supported by the high maximum mass predicted by a quark equation of state (G3 EOS-q1 and EOS-q2) and is also consistent with the analysis of observed data, as evident from Fig. \ref{fig4} and \ref{fig5}.

Another important macroscopic characteristic that can provide insights into the internal structure of these compact objects and put constraints on the underlying EOS is tidal deformability. Tidal deformability is a significant measure of the extent of deformation caused by tidal forces in compact stars, which can serve as a crucial constraint for determining the transition density for quark matter within neutron stars. The magnitude of tidal deformability in binary neutron stars directly impacts the intensity of the gravitational radiation generated during their inspiral phase, making it an essential factor to take into account \citep{PhysRevD.102.084058}. We examined the impacts on the relationship between dimensionless tidal deformability and the maximum mass of the star using the pure baryonic and quarkyonic EOSs developed previously. The dimensionless tidal deformability, which is a measure of a neutron star's deformability in response to a gravitational tidal field, is defined by the equation $\tilde{\Lambda}=\lambda/M^5$. Here, $\lambda$ represents the tidal deformability and is related to the induced quadruple deformation $Q_{ij}$ caused by the tidal field $\epsilon_{ij}$, as expressed by the equation $Q_{ij}=-\lambda\epsilon_{ij}$ \citep{PhysRevD.77.021502, Chatziioannou2020}. The tidal deformability, $\lambda$, can be mathematically represented using the dimensionless quadrupole tidal Love number, $k_2$, and the radius of the star as $\lambda = \frac{2}{3} k_{2} R^{5}$. The tidal love number, $k_2$, is intricately linked to the structure and composition of each individual star. For a more comprehensive understanding of this parameter, one can refer to sources \citep{PhysRevD.81.123016, PhysRevD.85.123007, PhysRevC.95.015801, PhysRevD.105.123010} that provide detailed mathematical expressions and discussions on how to delve into the complexities of this parameter and calculate the tidal love number. 
%%%%%%%%%%%%%%%%%%%%%%%%%%%%%%%%%%%%%%%%%%%%%%%%%%%%%%%%%%
%%%%%%%%%%%%%%%%%%%%%%%%%%%%%%%%%%%%%%%%%%%%%%%%%%%%%%%%%%
The dimensionless tidal deformability ($\tilde{\Lambda}$) values, which have been calculated for all the considered EOSs of both the RMF parameter sets, are illustrated in Fig. \ref{fig5}. This figure also displays the analysis of gravitational observational constraints, alongside the corresponding credible levels. It has been noted that the tidal deformability of a neutron star displays a sharp and steady decrease as its gravitational mass increases. After comparing the outcomes of both RMF parameter sets, we noticed that the pure baryonic G3 parameter set, which is known for its stiffer equation of state, produced significantly higher values of $\tilde{\Lambda}$ as compared to the pure baryonic FSUGold parameter set. The existence of quarks within the core of a star leads to a higher dimensionless tidal deformability for a neutron star of a specific mass, causing the star to be less compact compared to its baryonic counterpart. We have displayed the posterior distribution data on $\tilde{\Lambda}$ in Fig. \ref{fig5} for the more massive component in the GW170817 event at $68$ and $90\%$ credible intervals (CI), as well as for the binary merger components in the GW190425 gravitational wave event ($68\%$ CI). This data has been provided by the Laser Interferometer Gravitational-Wave Observatory and is publicly accessible \footnote{\tt https://dcc.ligo.org/LIGO-P1800115/public} \footnote{ https://dcc.ligo.org/LIGO-P2000026/public}. The posterior distribution data obtained from observing the GW170817 and GW190425 events provides a broad range for lambda that is consistent with almost all the considered scenarios of baryonic and quarkyonic stars. The lower and upper limits for the canonical star tidal deformability of the secondary component in the GW190814 merger event are also indicated by hatched line denoting the extreme values ($\tilde{\Lambda} = 616^{+273}_{-158}$) \citep{Abbott2020}. Similar to the M-R profile, the computed curve for the tidal deformability of a canonical star with the pure baryonic equation of state does not conform to the constraint imposed by the GW190814 merger event. However, the G3 quarkyonic EOS-q2 and EOS-q3 agree well with the restriction set by the GW190814 merger event. This suggests that the secondary component of the binary system may be a quarkyonic star, and provides compelling evidence to support this possibility.
\section{Binary NS Merger}

%\textcolor{red}{Shamim Please note the name of EOS. \\ EOS-FSUGold or EOS-G3 are "Pure Baryonic FSUGold" and "Pure Baryonic G3". These are for pure baryonic case without any quark. \\
%$FSUGold_{q}$ and $G3_{q}$ are EOS-q2 case ($n_{t} = 0.3 fm^{-3}, \Lambda = 1400$). You can assign it FSUGold EOS q-2 or EOS q-2 for FSUGold. Same for G3. 

In this section, we have discussed inferences extracted from full 3D numerical relativity simulations of binary NS merger (BNSM) of equal mass binaries. The configurations constructed using the pure baryonic (PB) EOSs --- FSUGold and G3 (labelled as FSUGold PB and G3 PB) are compared with the configurations from the EOSs --- FSUGold EOS-q2 and G3 EOS-q2. We studied the merger of a low-mass binary (1.2v1.2 $M_\odot$) and a high-mass binary (1.4v1.4 $M_\odot$) for both EOS cases.

\subsection{\label{sec:formalism}Formalism and Numerical Setup}
In 3+1 formalism~\citep{adm}, the Einstein field equations are cast into 3+1 splitting where the 4-dimensional spacetime is foliated into sequences of spacelike 3-dimensional hypersurfaces. The line element in this formalism is given by,
\begin{eqnarray}
	ds^2=-\alpha^2 dt^2+\gamma_{ij}\left(dx^i+\beta^idt\right)\left(dx^j+\beta^jdt\right),
\end{eqnarray}
where lapse function $\alpha$ and shift vector $\beta^i$ are the gauge variables, and $\gamma_{ij}$ is the spatial 3-metric induced on each hypersurface which is connected by timelike normal vectors,
\begin{eqnarray}
    n^\mu=\frac{1}{\alpha}\left(1,-\beta^i\right).
\end{eqnarray}
Here, the extrinsic curvature of the embedded hypersurface is defined as,
\begin{eqnarray}
    \kappa_{ij}=-\frac{1}{2}\mathcal{L}_\mathbf{n}\gamma_{ij},
\end{eqnarray}
where $\mathcal{L}_\mathbf{n}$ is the Lie derivative along $n^\mu$. Using this formalism, the Einstein field equations separate into set of constraint and evolution equations. It splits the stress-energy tensor in the following way:
\begin{small}\begin{eqnarray}
\bar{\rho}=n_{a} n_{b} T^{a b},\quad S^{i}=-\gamma^{i j}n^{a} T_{a j},\quad S_{i j}=\gamma_{i a} \gamma_{j b} T^{a b},
\end{eqnarray}\end{small}
\noindent where  $\bar{\rho}$ is the total energy density, $S^i$ is the total momentum density, and $S_{ij}$ spatial stress, measured by the normal (or Eulerian) observer $n^a$. A detailed discussion can be found in the following sources ---~\citep{adm10,adm11,adm12,adm13}.

For the evolution of BNSM systems, we use the \textsc{Einstein Toolkit}~\citep{ET2,ETextra1,ETextra2,ETextra3}. It is an open-source, community-driven computational infrastructure that is dedicated to simulate relativistic astrophysical systems. It is based on the \textsc{Cactus Computational Toolkit}~\citep{cactus}, a software framework designed with \textsc{Carpet} adaptive mesh refinement (AMR)~\citep{carpet1,amr} driver for high performance computing. The implementation of spacetime evolution is carried out by the \textsc{McLachlan} code~\citep{mclachlan1,mclachlan2}. It implements the Baumgarte-Shapiro-Shibata-Nakamura-Oohara-Kojima (BSSNOK)  formalism~\citep{bssn1,bssn2,bssn3,bssn4,bssn5} for evolution of the spacetime variables. Here, $\gamma_{ij}$ is conformally transformed as,
\begin{eqnarray}
    \Phi=\frac{1}{12}\mathrm{log}\left(\mathrm{det}\gamma_{ij}\right),\quad\Tilde{\gamma}_{ij}=e^{-4\Phi}\gamma_{ij}
\end{eqnarray}
where, $\Phi$ is the logarithmic conformal factor  and $\Tilde{\gamma}_{ij}$ is the conformal metric  (constrained by $\mathrm{det}\Tilde{\gamma}_{ij}=1$). These are the new variables alongside the trace of extrinsic curvature $\kappa$, the conformal trace free extrinsic curvature $\Tilde{A}_{ij}$ and the conformal connection functions $\Tilde{\Gamma}^i$, which are defined as,
\begin{small}
\begin{eqnarray}
\kappa=g^{ij}\kappa_{ij},~~\Tilde{A}_{ij}=e^{-4\Phi}\left(\kappa_{ij}-\frac{1}{3}g_{ij}\kappa\right),~~\Tilde{\Gamma}^i=\Tilde{\gamma}^{jk}\Tilde{\Gamma}^i_{jk}
\end{eqnarray}
\end{small}
\noindent to be evolved using the fourth order finite-differencing method. The gauge functions are determined using $1+\mathrm{log}$ slicing (for lapse function) and $\Gamma$-driver shift (for shift vectors) condition~\citep{gauge}. During the evolution, a Sommerfeld-type radiative boundary condition~\citep{sommerfeld} is applied to all the components of the evolved BSSNOK variables, and to discard the high-frequency noise, a fifth order Kreiss-Oliger dissipation term is added using the module-\textsc{Dissipation}.

The stress-energy tensor is given as,
\begin{eqnarray}
    T^{\mu\nu}=Pg^{\mu\nu}+\rho hu^\mu u^\nu,
\end{eqnarray}
where $P$ is the proper gas pressure, $\rho$ is the rest-mass density, $h$ is specific enthalpy, and $u^{\mu}$ is the 4-velocity of the fluid flow. The general relativistic (ideal) hydrodynamic (GRHD) equations are given by:
\begin{eqnarray}
    \nabla_\mu(\rho u^\mu)=0,\quad  \nabla_\mu(T^{\mu\nu})=0,
\end{eqnarray} where, $\nabla_\mu$ is the covariant derivative related to $g^{\mu\nu}$. These are the conservation equations of baryonic number and energy-momentum, which are closed by the equation of state of matter (briefly discussed in Section~\ref{EoS}). To include these EOSs in merger simulations, we mimic them using piece-wise polytrope fitting~\citep{adm13,ppeos}. The process is to first break a tabulated EOS into $N$ pieces (typically $4\sim8$) of density ranges. For each piece $i$ (range $\rho_{i}\leq\rho<\rho_{i+1}$),
\begin{eqnarray}
    P_\mathrm{c}=K_i\rho^{\Gamma_i},
\end{eqnarray}
where ($K_i,\Gamma_i$) are the $i^\mathrm{th}$ polytropic constant and polytropic exponent respectively. These pieces are matched at the boundary to ensure the smoothness of the EOS. These EOSs are supplemented by an ideal-fluid thermal component~\citep{gth2} which accounts for shock heating in the system that dissipates the kinetic energy into the internal energy,
\begin{eqnarray}
    P(\rho,\epsilon)=P_\mathrm{c}(\rho)+P_\mathrm{th}(\rho,\epsilon)=K_i{\rho}^{\Gamma_i}+\Gamma_\mathrm{th}\rho(\epsilon-\epsilon_\mathrm{c}),
\end{eqnarray}
where $\epsilon$ is the specific internal energy, and $\epsilon_c$ is given by,
\begin{eqnarray}
    \epsilon_\mathrm{c}=\epsilon_i+\frac{K_i}{\Gamma_i-1}\rho^{\Gamma_i-1}.
\end{eqnarray}
The thermal component $\Gamma_\mathrm{th}$ is set to 1.8 \citep{gth}. We used the \textsc{IllinoisGRMHD} code~\citep{illinois1,illinois2} for solving the GRMHD equations in 3+1 formalism, which are defined in a conservative form, and the flux terms are calculated using the second-order finite-volume high-resolution shock capturing (HRSC) scheme~\citep{adm13}, ensuring the Rankine-Hugoniot shock jump conditions. A third-order accurate piece-wise parabolic method (PPM)~\citep{ppm} is used for the reconstruction step. The standard Harten-Lax-van Leer-Einfeldt (HLLE) approximate Riemann solver~\citep{hlle1,hlle2} is applied. The method of lines module-\textsc{MoL} takes the time derivatives of the evolved GRMHD variables and integrates them forward in time using the Runge-Kutta fourth-order (RK4) scheme~\citep{rk1,rk2}. A two-dimensional Newton-Raphson solver is employed to compute the primitive variables from the conservative variables~\citep{p2c1,p2c2}.

The initial configuration data for our simulations are generated using the \texttt{Bin Star} code from the \textsc{Lorene} library~\citep{lorene2}. The \textsc{Lorene} uses multi-domain spectral methods to solve the partial differential equations. These data are obtained using the assumptions of quasi-circular equilibrium in the coalescence state and conformally flat metric to solve the conformal thin-sandwich equations~\citep{lorene3}. The grid and iteration parameters for our initial configurations were set identical to the version available at the \textsc{Subversion} repository server of the Gravitational Physics Group at the Parma University~\citep{parma1}. We have consistently set the initial physical separation between the stars to be $40~\mathrm{km}$ with irrotationality of the fluid flow.

For computing the gravitational waveforms from our simulations, we extract the Weyl scalar (particularly $\Psi_4$) from the simulations using the Newman-Penrose formalism~\citep{np}. We analysed the dominant mode $l=m=2$ of ${h}$ strain at 100 Mpc. We have set the \textit{merger time} at the point where the amplitude of the strain $|h^{22}|$ is maximum. The \textit{instantaneous frequency} is calculated by 
\begin{eqnarray}
 f_\mathrm{GW}=(1/2\pi)(d\phi/dt)   
\end{eqnarray}
We calculated the Power Spectral Density (PSD) of the gravitational wave (GW) amplitude ($\tilde{h}$), given as,
\begin{eqnarray}
    2\tilde{h}^2= |\tilde{h}_+|^2+|\tilde{h}_\times|^2 
\end{eqnarray}
where, $\tilde{h}_{+,\times}(f)$ are the Fourier transforms of ${h_{+,\times}(t)}$. A detailed discussion can be found in the following sources --- \citep{gw_ana,sam}. We used {\textsc{Kuibit}~\citep{kuibit1}} for GW data handling.

\subsection{Results}
\begin{figure}
    \centering
    \includegraphics[scale=0.58]{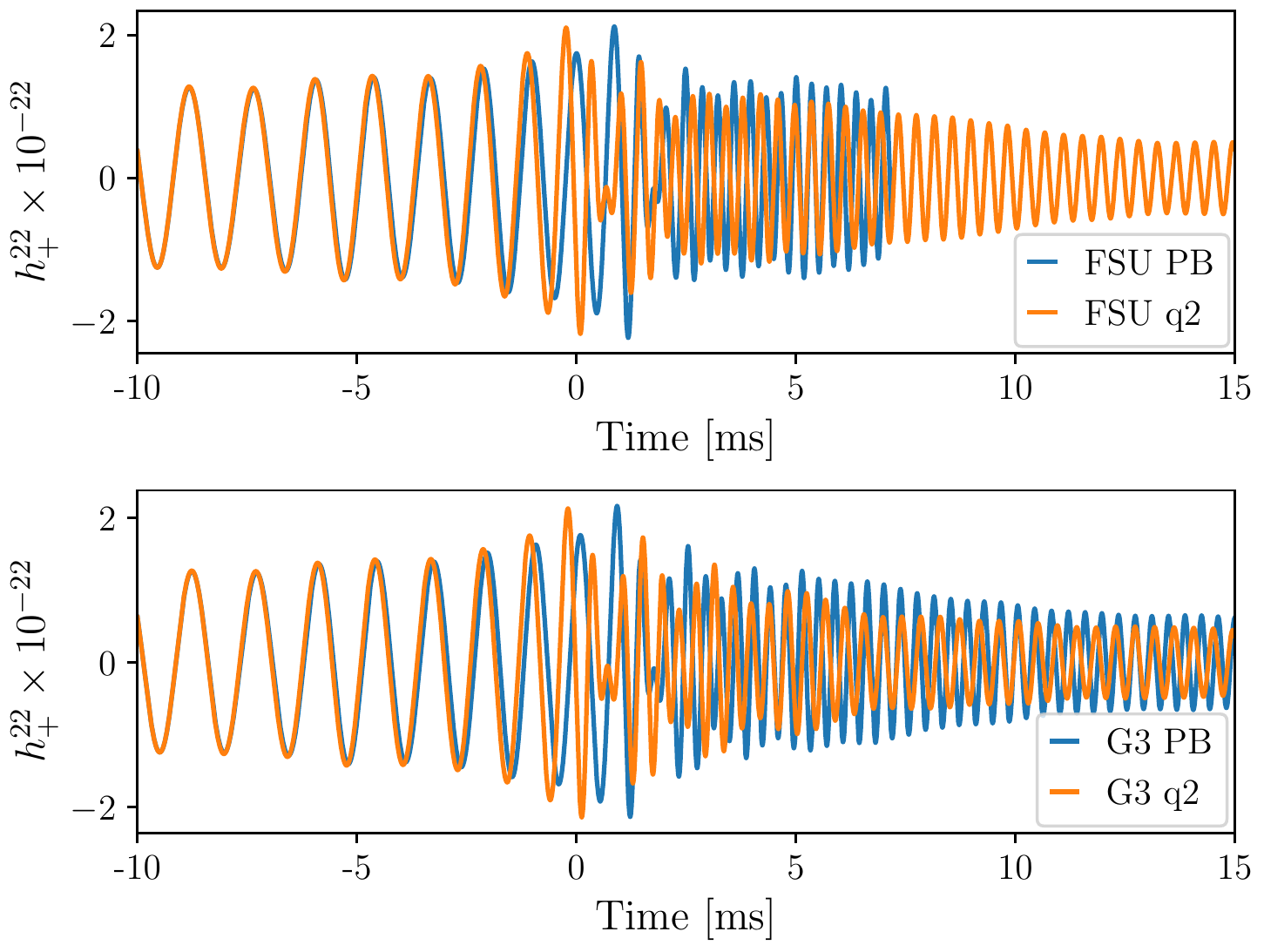}
    \caption{The $h^{22}_+$ polarisation of the GW signal extracted at 100 Mpc for 1.2v1.2 merger case. [Top] The GW signal extracted from EOS FSUGold (pure baryonic and quarkyonic). [Bottom] The GW signal extracted from EOS G3 (pure baryonic and quarkyonic).}
    \label{fig:et_gw}
\end{figure}

We simulated two mass equal binary configurations --- 1.2v1.2 $M_\odot$ and 1.4v1.4 $M_\odot$ for each EOS case. In Fig.~\ref{fig:et_gw} and \ref{fig:et_rhomax}, we have set $t=0$ as a reference of \textit{merger time} computed from the simulations constructed using quarkyonic EOS. In Fig.~\ref{fig:et_gw}, we plotted the $h_+^{22}$ polarisation of the GW signal for 1.2v1.2 $M_\odot$ merger case. In the top panel, we compared the GW signals from simulations of BNSM using Pure Baryonic EOS FSUGold and FSUGold EOS-q2. In the bottom panel, a similar comparison is made for Pure Baryonic EOS G3 and G3 EOS-q2.
In both EOS merger cases (FSUGold and G3), we observe that the GW signal is similar till 5~ms before $t=0$. The signal starts to deviate during the final cycles of the inspiral phase, which appears due to the difference in the tidal deformability of the 1.2 $M_\odot$ constructed using pure baryonic EOS and quarkyonic EOS, as discussed in Sec.~\ref{Static}. We also note that quarkyonic EOS advances the merger time (at the scale of 1~ms) with respect to pure baryonic EOS. In both panels (top and bottom), we observe that the post-merger signals from the quarkyonic EOS mergers deviate significantly with respect to their own pure baryonic EOS.

\begin{figure}
    \centering
    \includegraphics[scale=0.58]{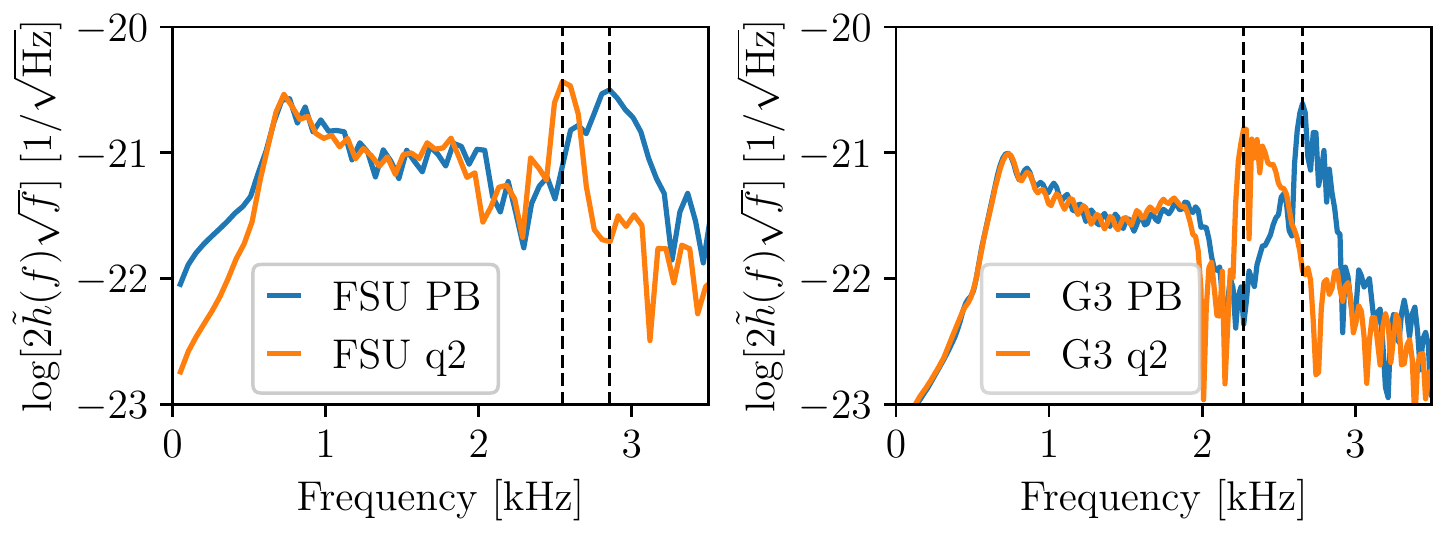}
    \caption{PSDs of GW signals plotted in Fig.~\ref{fig:et_gw}, marked with the $f_2$ frequencies. [Left] The PSD of the GW signal extracted from EOS FSUGold (pure baryonic and quarkyonic). [Right] The PSD of GW signal extracted from EOS G3 (pure baryonic and quarkyonic).}
    \label{fig:et_psd}
\end{figure}

\begin{figure}
    \centering
    \includegraphics[scale=0.6]{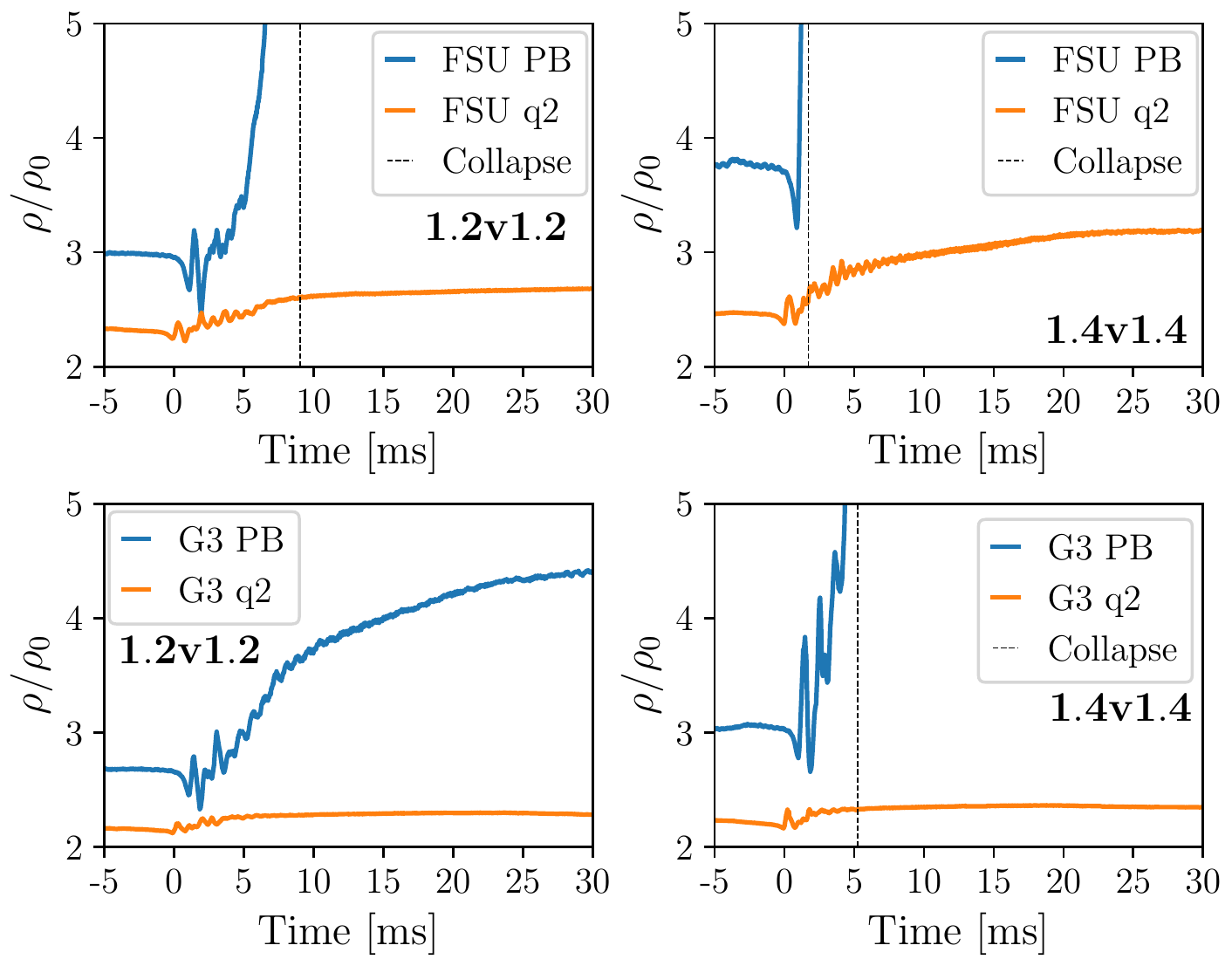}
    \caption{Evolution of maximum density ($\rho_{\mathrm{max}}$). Collapse times are marked for the cases where compact objects collapsed into BHs. [Top] Evolution of $\rho_{\mathrm{max}}$ from EOS FSUGold (pure baryonic and quarkyonic). [Bottom] The GW signal extracted from EOS G3 (pure baryonic and quarkyonic). [Left] Evolution of $\rho_{\mathrm{max}}$ for 1.2v1.2 $M_\odot$ merger. [Right] Evolution of $\rho_{\mathrm{max}}$ for 1.4v1.4 $M_\odot$ merger.}
    \label{fig:et_rhomax}
\end{figure}

This difference is also reflected in their PSDs, as observed in Fig.~\ref{fig:et_psd}. We have marked the $f_2$ frequencies in the PSD for both cases. These frequencies are twice the rotational frequency of the bar deformations of the hypermassive remnants. A detailed discussion about the spectral properties of the hypermassive remnants in numerical relativity simulations can be found in Ref.~\cite{gw_ana}. For the case of Pure Baryonic EOS FSUGold and FSUGold EOS-q2 (left panel), the $f_2$ frequencies are 2.84~kHz and 2.58~kHz, respectively. For the case of Pure Baryonic EOS G3 and G3 EOS-q2 (right panel), the $f_2$ frequencies are 2.65~kHz and 2.27~kHz, respectively. In both cases, we observed slower rotational frequencies of the hypermassive remnant for the quarkyonic EOS when compared to their pure baryonic EOS.

In Fig.~\ref{fig:et_rhomax}, we plotted the maximum density ($\rho_{\mathrm{max}}$) evolution for all BNSM simulations. The densities are given in terms of the nuclear saturation density $\rho_0$ ($2.51 \times 10^{14}~\mathrm{g/cm^3}$, \cite{glen}). For scenarios of core collapsing in black hole (BH), we identify the onset of collapse when the maximum density in the simulation instantaneously peaks ($30\sim 100\rho_0$ within $\sim 0.5$ ms) and mark the {\it collapse time} when $\rho_{\mathrm{max}}$ goes beyond $25\rho_0$. 

In Fig.~\ref{fig:et_rhomax} (top left), for the case of 1.2v1.2 $M_\odot$ merger with Pure Baryonic EOS FSUGold, the merger remnant collapses into BH within 10~ms. The characteristic of this particular merger collapsing into BH can be explained by the fact the maximum mass of NS that can be constructed using Pure Baryonic EOS FSUGold is $\sim1.75~M_\odot$, as observed from its M-R curves in Fig.~\ref{fig4}. In contrast to the previous case, the merger remnant constructed using FSUGold EOS-q2 becomes a hypermassive NS (HMNS). The maximum density evolution saturates at $\sim2.7\rho_0$ and does not collapse till 45~ms of evolution, as observed in Fig.~\ref{fig:et_rhomax} (top-right). It confirms the stiffer nature of FSUGold EOS-q2 with respect to Pure Baryonic EOS FSUGold. For 1.4v1.4 $M_\odot$ merger case, which is constructed using Pure Baryonic EOS FSUGold, collapses immediately after the first contact between the two NSs. However, from the merger constructed using FSUGold EOS-q2, the merger remnant survives to form an HMNS, where the maximum density saturates at $\sim3.2\rho_0$.

For the case of 1.2v1.2 $M_\odot$ mergers constructed using Pure Baryonic EOS G3 and G3 EOS-q2, both the merger remnants form HMNS. In Fig.~\ref{fig:et_rhomax} (bottom left), we observe that the maximum density evolution of the merger remnant formed using Pure Baryonic EOS G3 saturates at much higher density than the one formed using G3 EOS-q2. For 1.4v1.4 $M_\odot$ mergers constructed using Pure Baryonic EOS G3, the hypermassive remnant collapses into BH within 5~ms of post-merger evolution. However, the merger remnant in the case of G3 EOS-q2 survives to form an HMNS. It does not collapse till 45~ms of evolution, as observed in Fig.~\ref{fig:et_rhomax} (bottom-right). The maximum density evolution of this HMNS saturates at $\sim2.4\rho_0$. It hints at the stiffer nature of G3 EOS-q2 with respect to Pure Baryonic EOS G3.

In both EOS cases (FSUGold and G3), it is observed that quarkyonic EOS being stiffer in nature with respect to pure baryonic EOS, does not favour the core collapse of the hypermassive remnant for intermediate mass merger (1.4v1.4 $M_\odot$) even when its own pure baryonic EOS is quite softer (like EOS G3), favouring core collapse to BH scenario.
%%%%%%%%%%%%%%%%%%%%%%%%%%%%%%%%%%%%%%%
%%%%%%%%%%%%%%%%%%%%%%%%%%%%%%%%%%%%%%%
%%%%%%%%%%%%%%%%%%%%%%%%%%%%%%%%%%%%%%%
%%%%%%%%%%%%%%%%%%%%%%%%%%%%%%%%%%%%%%%
%%%%%%%%%%%%%%%%%%%%%%%%%%%%%%%%%%%%%%%
%%%%%%%%%%%%%%%%%%%%%%%%%%%%%%%%%%%%%%%
%%%%%%%%%%%%%%%%%%%%%%%%%%%%%%%%%%%%%%%
\section{Conclusions} In conclusion, we examined the impact of incorporating quarkyonic matter in neutron stars by formulating the quarkyonic star equation of state using the relativistic mean field approach and studying the static and merger properties. The quarkyonic model employs two free parameters: the transition density and QCD confinement scale. The former affects the speed of sound and makes the equation of state stiffer, while the latter calibrates the maximum mass of a neutron star. Interestingly, speed of sound for all quarkyonic EOS converges to the conformal limit $c/\sqrt3$ due to the formation of deconfined relativistic quarks at high density, which impressively shows a robust agreement with speed of sound bounds inferred from gravitational wave data. The results of theoretical simulation also suggests that the quarkyonic EOS is the only one that can accurately predict the masses of massive neutron stars that are larger than $2M_\odot$ while still respecting the observed conformal limit for the speed of sound at high densities. Therefore, it is feasible that massive neutron stars like the "black widow" pulsar PSR J0952-0607 and the secondary component in the GW190814 event are actually quarkyonic stars. The concurrence with tidal deformability estimates derived from the GW190814 event along with GW170817 and GW190425 events with the predicted values correspond to quarkyonic EOS further validates the quarkyonic model as a consistent alternative approach for parameterized description of ultra dense matter.\\

%We found that the quarkyonic model provides a more consistent approach for describing highly dense matter by introducing two free parameters, the transition density and QCD confinement scale. The inclusion of the transition density leads to a stiffer equation of state, while the QCD confinement scale affects the maximum allowable speed of sound value, which can precisely calibrate the predicted maximum mass. Interestingly, we find that speed of sound for all quarkyonic EOS converges to conformal limit $c/\sqrt{3}$ at high density due to deconfined relativistic quarks  which impressively shows high degree of coherence with speed of sound bounds inferred from gravitational wave data. We also observe that only the quarkyonic EOS can predict massive neutron star masses greater than $2M_\odot$, even reaching heavier mass range $2.6 M_{\odot}$ which is consistent with observational data. The agreement with tidal deformability estimates derived from the GW170817 and GW190425 events further validates the accuracy of the quarkyonic model in characterizing the underlying quarkyonic equation of state.\\
The post-merger dynamics of binary neutron star merger were studied from the gravitational wave analysis and density evolution using the numerical relativity simulations of two equal-mass configurations (1.2v1.2 $M_\odot$ and 1.4v1.4 $M_\odot$). Results showed that the quarkyonic equation of state disfavors the scenarios of the merger remnant core collapsing into a black hole. It indicates the impact of stiffening due to the crossover transition between hadrons and quarks. The occurrence of quarkyonic matter can also advance the merger time at a scale of 1~ms. Furthermore, a distinct decrease in $f_2$ frequency was observed for the quarkyonic EOS, which indicates at the lower rotational frequency of the bar deformations of the hypermassive remnants. It may differentiate between quarkyonic and pure baryonic matter, indicating the presence of quarks inside the neutron star.

%Several unexplored possibilities and aspects of the model must be considered to have a comprehensive understanding of neutron star. As more data becomes available from gravitational wave and multimessenger astronomy, the bounds on speed of sound, mass-radius, and tidal deformability are expected to become more stringent and improvements to the current model may be necessary to accommodate these bounds. Also, inclusion of other components like strange baryons , dark matter should also be considered for a more complete and general treatment of neutron star. Taking account the rotation of neutron stars can significantly alter nature and structure of  matter and may play a vital role in driving phase transitions. These avenues of research have the potential to deepen our understanding of ultra-dense matter inside neutron stars and are subject of future investigation 

As more data becomes available from observation of gravitational waves and multi-messenger astronomy, the bounds on the speed of sound, mass-radius, and tidal deformability are expected to become more stringent, and improvements to the current model may be necessary to meet these constraints. Also, including the strange quark in our study may reveal the formation of coulomb lattices that could explain glitches in pulsars and determine the thermal and transport properties of neutron stars. Other constituents such as strange baryons, and dark matter should also be considered for a more comprehensive understanding of these compact objects. These avenues of research have the potential to deepen our understanding of ultra-dense matter inside neutron stars and are the subject of future investigation.

\section*{acknowledgments}
SH and RM thank IISER Bhopal for providing the infrastructure facilities for the numerical relativity simulations.
%%%%%%%%%%%%%%%
\section*{Data Availability}
Sharing of the data generated in this article will be facilitated upon request to the corresponding author on reasonable grounds.
%%%%%%%%%%%%%%
\bibliography{Quarkyonic}

\begin{thebibliography}{}
\makeatletter
\relax
\def\mn@urlcharsother{\let\do\@makeother \do\$\do\&\do\#\do\^\do\_\do\%\do\~}
\def\mn@doi{\begingroup\mn@urlcharsother \@ifnextchar [ {\mn@doi@}
  {\mn@doi@[]}}
\def\mn@doi@[#1]#2{\def\@tempa{#1}\ifx\@tempa\@empty \href
  {http://dx.doi.org/#2} {doi:#2}\else \href {http://dx.doi.org/#2} {#1}\fi
  \endgroup}
\def\mn@eprint#1#2{\mn@eprint@#1:#2::\@nil}
\def\mn@eprint@arXiv#1{\href {http://arxiv.org/abs/#1} {{\tt arXiv:#1}}}
\def\mn@eprint@dblp#1{\href {http://dblp.uni-trier.de/rec/bibtex/#1.xml}
  {dblp:#1}}
\def\mn@eprint@#1:#2:#3:#4\@nil{\def\@tempa {#1}\def\@tempb {#2}\def\@tempc
  {#3}\ifx \@tempc \@empty \let \@tempc \@tempb \let \@tempb \@tempa \fi \ifx
  \@tempb \@empty \def\@tempb {arXiv}\fi \@ifundefined
  {mn@eprint@\@tempb}{\@tempb:\@tempc}{\expandafter \expandafter \csname
  mn@eprint@\@tempb\endcsname \expandafter{\@tempc}}}

\bibitem[\protect\citeauthoryear{{Abbott} et~al.,}{{Abbott}
  et~al.}{2017}]{2017PhRvL.119p1101A}
{Abbott} B.~P.,  et~al., 2017, \mn@doi [\prl] {10.1103/PhysRevLett.119.161101},
  \href {https://ui.adsabs.harvard.edu/abs/2017PhRvL.119p1101A} {119, 161101}

\bibitem[\protect\citeauthoryear{{Abbott} et~al.,}{{Abbott}
  et~al.}{2020a}]{2020ApJ...892L...3A}
{Abbott} B.~P.,  et~al., 2020a, \mn@doi [\apjl] {10.3847/2041-8213/ab75f5},
  \href {https://ui.adsabs.harvard.edu/abs/2020ApJ...892L...3A} {892, L3}

\bibitem[\protect\citeauthoryear{Abbott et~al.}{Abbott
  et~al.}{2020b}]{Abbott2020}
Abbott R.,  et~al., 2020b, \mn@doi [The Astrophysical Journal Letters]
  {10.3847/2041-8213/ab960f}, 896, L44

\bibitem[\protect\citeauthoryear{Alcubierre, Br\"ugmann, Dramlitsch, Font,
  Papadopoulos, Seidel, Stergioulas  \& Takahashi}{Alcubierre
  et~al.}{2000a}]{bssn4}
Alcubierre M.,  Br\"ugmann B.,  Dramlitsch T.,  Font J.~A.,  Papadopoulos P.,
  Seidel E.,  Stergioulas N.,   Takahashi R.,  2000a, \mn@doi [Phys. Rev. D]
  {10.1103/PhysRevD.62.044034}, 62, 044034

\bibitem[\protect\citeauthoryear{Alcubierre, Br\"ugmann, Dramlitsch, Font,
  Papadopoulos, Seidel, Stergioulas  \& Takahashi}{Alcubierre
  et~al.}{2000b}]{sommerfeld}
Alcubierre M.,  Br\"ugmann B.,  Dramlitsch T.,  Font J.~A.,  Papadopoulos P.,
  Seidel E.,  Stergioulas N.,   Takahashi R.,  2000b, \mn@doi [Phys. Rev. D]
  {10.1103/PhysRevD.62.044034}, 62, 044034

\bibitem[\protect\citeauthoryear{Alcubierre, Br\"ugmann, Diener, Koppitz,
  Pollney, Seidel  \& Takahashi}{Alcubierre et~al.}{2003a}]{bssn5}
Alcubierre M.,  Br\"ugmann B.,  Diener P.,  Koppitz M.,  Pollney D.,  Seidel
  E.,   Takahashi R.,  2003a, \mn@doi [Phys. Rev. D]
  {10.1103/PhysRevD.67.084023}, 67, 084023

\bibitem[\protect\citeauthoryear{Alcubierre, Br\"ugmann, Diener, Koppitz,
  Pollney, Seidel  \& Takahashi}{Alcubierre et~al.}{2003b}]{gauge}
Alcubierre M.,  Br\"ugmann B.,  Diener P.,  Koppitz M.,  Pollney D.,  Seidel
  E.,   Takahashi R.,  2003b, \mn@doi [Phys. Rev. D]
  {10.1103/PhysRevD.67.084023}, 67, 084023

\bibitem[\protect\citeauthoryear{Altiparmak, Ecker  \& Rezzolla}{Altiparmak
  et~al.}{2022}]{Altiparmak_2022}
Altiparmak S.,  Ecker C.,   Rezzolla L.,  2022, \mn@doi [The Astrophysical
  Journal Letters] {10.3847/2041-8213/ac9b2a}, 939, L34

\bibitem[\protect\citeauthoryear{Annala, Gorda, Kurkela  \& Vuorinen}{Annala
  et~al.}{2018}]{PhysRevLett.120.172703}
Annala E.,  Gorda T.,  Kurkela A.,   Vuorinen A.,  2018, \mn@doi [Phys. Rev.
  Lett.] {10.1103/PhysRevLett.120.172703}, 120, 172703

\bibitem[\protect\citeauthoryear{Arnowitt, Deser  \& Misner}{Arnowitt
  et~al.}{2008}]{adm}
Arnowitt R.,  Deser S.,   Misner C.~W.,  2008, \mn@doi [General Relativity and
  Gravitation] {10.1007/s10714-008-0661-1}, 40, 1997

\bibitem[\protect\citeauthoryear{Baumgarte \& Shapiro}{Baumgarte \&
  Shapiro}{1998}]{bssn3}
Baumgarte T.~W.,  Shapiro S.~L.,  1998, \mn@doi [Phys. Rev. D]
  {10.1103/PhysRevD.59.024007}, 59, 024007

\bibitem[\protect\citeauthoryear{Baumgarte \& Shapiro}{Baumgarte \&
  Shapiro}{2010}]{adm10}
Baumgarte T.~W.,  Shapiro S.~L.,  2010, Numerical Relativity: Solving
  Einstein's Equations on the Computer.
Cambridge University Press, \mn@doi{10.1017/CBO9781139193344}

\bibitem[\protect\citeauthoryear{{Beni\'{}c, Sanjin}, {Blaschke, David},
  {Alvarez-Castillo, David E.}, {Fischer, Tobias}  \& {Typel,
  Stefan}}{{Beni\'{}c, Sanjin} et~al.}{2015}]{refId0}
{Beni\'{}c, Sanjin} {Blaschke, David} {Alvarez-Castillo, David E.} {Fischer,
  Tobias}  {Typel, Stefan} 2015, \mn@doi [A\&A] {10.1051/0004-6361/201425318},
  577, A40

\bibitem[\protect\citeauthoryear{Biswal, Patra  \& Zhou}{Biswal
  et~al.}{2019}]{Biswal_2019}
Biswal S.~K.,  Patra S.~K.,   Zhou S.-G.,  2019, \mn@doi [The Astrophysical
  Journal] {10.3847/1538-4357/ab43c5}, 885, 25

\bibitem[\protect\citeauthoryear{Boguta \& Bodmer}{Boguta \&
  Bodmer}{1977}]{BOGUTA1977413}
Boguta J.,  Bodmer A.,  1977, \mn@doi [Nuclear Physics A]
  {https://doi.org/10.1016/0375-9474(77)90626-1}, 292, 413

\bibitem[\protect\citeauthoryear{{Bozzola}}{{Bozzola}}{2021}]{kuibit1}
{Bozzola} G.,  2021, \mn@doi [The Journal of Open Source Software]
  {10.21105/joss.03099}, \href
  {https://ui.adsabs.harvard.edu/abs/2021JOSS....6.3099B} {6, 3099}

\bibitem[\protect\citeauthoryear{Brown, Diener, Sarbach, Schnetter  \&
  Tiglio}{Brown et~al.}{2009}]{mclachlan2}
Brown D.,  Diener P.,  Sarbach O.,  Schnetter E.,   Tiglio M.,  2009, \mn@doi
  [Phys. Rev. D] {10.1103/PhysRevD.79.044023}, 79, 044023

\bibitem[\protect\citeauthoryear{Bult et~al.,}{Bult et~al.}{2019}]{Bult_2019}
Bult P.,  et~al., 2019, \mn@doi [The Astrophysical Journal Letters]
  {10.3847/2041-8213/ab4ae1}, 885, L1

\bibitem[\protect\citeauthoryear{Bunta \& Gmuca}{Bunta \&
  Gmuca}{2003}]{PhysRevC.68.054318}
Bunta J. K. c.~v.,  Gmuca i. c.~v.,  2003, \mn@doi [Phys. Rev. C]
  {10.1103/PhysRevC.68.054318}, 68, 054318

\bibitem[\protect\citeauthoryear{Bunta \& Gmuca}{Bunta \&
  Gmuca}{2004}]{PhysRevC.70.054309}
Bunta J. K. c.~v.,  Gmuca i. c.~v.,  2004, \mn@doi [Phys. Rev. C]
  {10.1103/PhysRevC.70.054309}, 70, 054309

\bibitem[\protect\citeauthoryear{Cai \& Chen}{Cai \&
  Chen}{2012}]{PhysRevC.85.024302}
Cai B.-J.,  Chen L.-W.,  2012, \mn@doi [Phys. Rev. C]
  {10.1103/PhysRevC.85.024302}, 85, 024302

\bibitem[\protect\citeauthoryear{Chatziioannou}{Chatziioannou}{2020}]{Chatziioannou2020}
Chatziioannou K.,  2020, \mn@doi [General Relativity and Gravitation]
  {10.1007/s10714-020-02754-3}, 52, 109

\bibitem[\protect\citeauthoryear{Colella \& Woodward}{Colella \&
  Woodward}{1984}]{ppm}
Colella P.,  Woodward P.~R.,  1984, \mn@doi [Journal of Computational Physics]
  {https://doi.org/10.1016/0021-9991(84)90143-8}, 54, 174

\bibitem[\protect\citeauthoryear{Dadi}{Dadi}{2010}]{PhysRevC.82.025203}
Dadi A. b.~A.,  2010, \mn@doi [Phys. Rev. C] {10.1103/PhysRevC.82.025203}, 82,
  025203

\bibitem[\protect\citeauthoryear{Damour, Nagar  \& Villain}{Damour
  et~al.}{2012}]{PhysRevD.85.123007}
Damour T.,  Nagar A.,   Villain L.,  2012, \mn@doi [Phys. Rev. D]
  {10.1103/PhysRevD.85.123007}, 85, 123007

\bibitem[\protect\citeauthoryear{Das, Kumar, Biswal  \& Patra}{Das
  et~al.}{2021a}]{PhysRevD.104.123006}
Das H.~C.,  Kumar A.,  Biswal S.~K.,   Patra S.~K.,  2021a, \mn@doi [Phys. Rev.
  D] {10.1103/PhysRevD.104.123006}, 104, 123006

\bibitem[\protect\citeauthoryear{Das, Kumar  \& Patra}{Das
  et~al.}{2021b}]{10.1093/mnras/stab2387}
Das H.~C.,  Kumar A.,   Patra S.~K.,  2021b, \mn@doi [Monthly Notices of the
  Royal Astronomical Society] {10.1093/mnras/stab2387}, 507, 4053

\bibitem[\protect\citeauthoryear{De~Pietri, Feo, Maione  \&
  L\"offler}{De~Pietri et~al.}{2016}]{parma1}
De~Pietri R.,  Feo A.,  Maione F.,   L\"offler F.,  2016, \mn@doi [Phys. Rev.
  D] {10.1103/PhysRevD.93.064047}, 93, 064047

\bibitem[\protect\citeauthoryear{Del~Estal, Centelles, Vi\~nas  \&
  Patra}{Del~Estal et~al.}{2001}]{PhysRevC.63.024314}
Del~Estal M.,  Centelles M.,  Vi\~nas X.,   Patra S.~K.,  2001, \mn@doi [Phys.
  Rev. C] {10.1103/PhysRevC.63.024314}, 63, 024314

\bibitem[\protect\citeauthoryear{{Del Zanna, L.}, {Bucciantini, N.}  \&
  {Londrillo, P.}}{{Del Zanna, L.} et~al.}{2003}]{illinois2}
{Del Zanna, L.} {Bucciantini, N.}  {Londrillo, P.} 2003, \mn@doi [A\&A]
  {10.1051/0004-6361:20021641}, 400, 397

\bibitem[\protect\citeauthoryear{Dexheimer \& Schramm}{Dexheimer \&
  Schramm}{2010}]{PhysRevC.81.045201}
Dexheimer V.~A.,  Schramm S.,  2010, \mn@doi [Phys. Rev. C]
  {10.1103/PhysRevC.81.045201}, 81, 045201

\bibitem[\protect\citeauthoryear{Diener, Dorband, Schnetter  \& Tiglio}{Diener
  et~al.}{2007}]{ETextra2}
Diener P.,  Dorband E.~N.,  Schnetter E.,   Tiglio M.,  2007, \mn@doi [Journal
  of Scientific Computing] {10.1007/s10915-006-9123-7}, 32, 109

\bibitem[\protect\citeauthoryear{Dreyer, Krishnan, Shoemaker  \&
  Schnetter}{Dreyer et~al.}{2003}]{ETextra3}
Dreyer O.,  Krishnan B.,  Shoemaker D.,   Schnetter E.,  2003, \mn@doi [Phys.
  Rev. D] {10.1103/PhysRevD.67.024018}, 67, 024018

\bibitem[\protect\citeauthoryear{Duarte, Hernandez-Ortiz  \& Jeong}{Duarte
  et~al.}{2020}]{PhysRevC.102.025203}
Duarte D.~C.,  Hernandez-Ortiz S.,   Jeong K.~S.,  2020, \mn@doi [Phys. Rev. C]
  {10.1103/PhysRevC.102.025203}, 102, 025203

\bibitem[\protect\citeauthoryear{Einfeldt}{Einfeldt}{1988}]{hlle2}
Einfeldt B.,  1988, \mn@doi [SIAM Journal on Numerical Analysis]
  {10.1137/0725021}, 25, 294

\bibitem[\protect\citeauthoryear{Etienne, Paschalidis, Haas, Mösta  \&
  Shapiro}{Etienne et~al.}{2015}]{illinois1}
Etienne Z.~B.,  Paschalidis V.,  Haas R.,  Mösta P.,   Shapiro S.~L.,  2015,
  \mn@doi [Classical and Quantum Gravity] {10.1088/0264-9381/32/17/175009}, 32,
  175009

\bibitem[\protect\citeauthoryear{Fattoyev, Horowitz, Piekarewicz  \&
  Shen}{Fattoyev et~al.}{2010}]{PhysRevC.82.055803}
Fattoyev F.~J.,  Horowitz C.~J.,  Piekarewicz J.,   Shen G.,  2010, \mn@doi
  [Phys. Rev. C] {10.1103/PhysRevC.82.055803}, 82, 055803

\bibitem[\protect\citeauthoryear{Fattoyev, Horowitz, Piekarewicz  \&
  Reed}{Fattoyev et~al.}{2020}]{PhysRevC.102.065805}
Fattoyev F.~J.,  Horowitz C.~J.,  Piekarewicz J.,   Reed B.,  2020, \mn@doi
  [Phys. Rev. C] {10.1103/PhysRevC.102.065805}, 102, 065805

\bibitem[\protect\citeauthoryear{Flanagan \& Hinderer}{Flanagan \&
  Hinderer}{2008}]{PhysRevD.77.021502}
Flanagan E.~E.,  Hinderer T.,  2008, \mn@doi [Phys. Rev. D]
  {10.1103/PhysRevD.77.021502}, 77, 021502

\bibitem[\protect\citeauthoryear{{Fonseca} et~al.,}{{Fonseca}
  et~al.}{2021}]{2021ApJ...915L..12F}
{Fonseca} E.,  et~al., 2021, \mn@doi [\apjl] {10.3847/2041-8213/ac03b8}, \href
  {https://ui.adsabs.harvard.edu/abs/2021ApJ...915L..12F} {915, L12}

\bibitem[\protect\citeauthoryear{Frauendiener}{Frauendiener}{2011}]{adm11}
Frauendiener J.,  2011, Miguel Alcubierre: Introduction to 3 + 1 numerical
  relativity.
Oxford University Press, \mn@doi{10.1007/s10714-011-1195-5}, \url
  {https://doi.org/10.1007/s10714-011-1195-5}

\bibitem[\protect\citeauthoryear{{Galloway}, {Goodwin}, {Heger}  \&
  {Johnston}}{{Galloway} et~al.}{2021}]{2021cosp...43E1213G}
{Galloway} D.,  {Goodwin} A.,  {Heger} A.,   {Johnston} Z.,  2021, in 43rd
  COSPAR Scientific Assembly. Held 28 January - 4 February. p.~1213

\bibitem[\protect\citeauthoryear{Gambhir \& Ring}{Gambhir \&
  Ring}{1989}]{Gambhir1989}
Gambhir Y.~K.,  Ring P.,  1989, \mn@doi [Pramana] {10.1007/BF02845972}, 32, 389

\bibitem[\protect\citeauthoryear{{Gautam} et~al.,}{{Gautam}
  et~al.}{2022}]{2022A&A...664A..54G}
{Gautam} T.,  et~al., 2022, \mn@doi [\aap] {10.1051/0004-6361/202243062}, \href
  {https://ui.adsabs.harvard.edu/abs/2022A&A...664A..54G} {664, A54}

\bibitem[\protect\citeauthoryear{{Glendenning}}{{Glendenning}}{1985}]{1985ApJ...293..470G}
{Glendenning} N.~K.,  1985, \mn@doi [\apj] {10.1086/163253}, \href
  {https://ui.adsabs.harvard.edu/abs/1985ApJ...293..470G} {293, 470}

\bibitem[\protect\citeauthoryear{Glendenning}{Glendenning}{1992}]{PhysRevD.46.1274}
Glendenning N.~K.,  1992, \mn@doi [Phys. Rev. D] {10.1103/PhysRevD.46.1274},
  46, 1274

\bibitem[\protect\citeauthoryear{Glendenning}{Glendenning}{1997}]{glen}
Glendenning N.~K.,  1997, Compact Stars.
Springer New York, NY, \mn@doi{10.1007/978-1-4684-0491-3}

\bibitem[\protect\citeauthoryear{Glendenning, Weber  \& Moszkowski}{Glendenning
  et~al.}{1992}]{PhysRevC.45.844}
Glendenning N.~K.,  Weber F.,   Moszkowski S.~A.,  1992, \mn@doi [Phys. Rev. C]
  {10.1103/PhysRevC.45.844}, 45, 844

\bibitem[\protect\citeauthoryear{Goodale, Allen, Lanfermann, Mass{\'o}, Radke,
  Seidel  \& Shalf}{Goodale et~al.}{2003}]{cactus}
Goodale T.,  Allen G.,  Lanfermann G.,  Mass{\'o} J.,  Radke T.,  Seidel E.,
  Shalf J.,  2003, in High Performance Computing for Computational Science ---
  VECPAR 2002. Springer Berlin Heidelberg, \mn@doi{10.1007/3-540-36569-9_13}

\bibitem[\protect\citeauthoryear{Gourgoulhon}{Gourgoulhon}{2012}]{adm12}
Gourgoulhon E.,  2012, 3+1 Formalism in General Relativity.
GSpringer Berlin, Heidelberg, \mn@doi{10.1007/978-3-642-24525-1}, \url
  {https://doi.org/10.1007/978-3-642-24525-1}

\bibitem[\protect\citeauthoryear{Gourgoulhon, Grandcl\'ement, Taniguchi, Marck
  \& Bonazzola}{Gourgoulhon et~al.}{2001}]{lorene2}
Gourgoulhon E.,  Grandcl\'ement P.,  Taniguchi K.,  Marck J.-A.,   Bonazzola
  S.,  2001, \mn@doi [Phys. Rev. D] {10.1103/PhysRevD.63.064029}, 63, 064029

\bibitem[\protect\citeauthoryear{Greif, Raaijmakers, Hebeler, Schwenk  \&
  Watts}{Greif et~al.}{2019}]{10.1093/mnras/stz654}
Greif S.~K.,  Raaijmakers G.,  Hebeler K.,  Schwenk A.,   Watts A.~L.,  2019,
  \mn@doi [Monthly Notices of the Royal Astronomical Society]
  {10.1093/mnras/stz654}, 485, 5363

\bibitem[\protect\citeauthoryear{Haque, Mallick  \& Thakur}{Haque
  et~al.}{2022}]{sam}
Haque S.,  Mallick R.,   Thakur S.~K.,  2022 (\mn@eprint {arXiv} {2207.14485})

\bibitem[\protect\citeauthoryear{{Harris} \& {Alford}}{{Harris} \&
  {Alford}}{2018}]{2018APS..APRS11009H}
{Harris} S.,  {Alford} M.,  2018, in APS April Meeting Abstracts. p. S11.009

\bibitem[\protect\citeauthoryear{Harten, Lax  \& Leer}{Harten
  et~al.}{1983}]{hlle1}
Harten A.,  Lax P.~D.,   Leer B.~v.,  1983, \mn@doi [SIAM Review]
  {10.1137/1025002}, 25, 35

\bibitem[\protect\citeauthoryear{Hinderer, Lackey, Lang  \& Read}{Hinderer
  et~al.}{2010}]{PhysRevD.81.123016}
Hinderer T.,  Lackey B.~D.,  Lang R.~N.,   Read J.~S.,  2010, \mn@doi [Phys.
  Rev. D] {10.1103/PhysRevD.81.123016}, 81, 123016

\bibitem[\protect\citeauthoryear{Hornick, Tolos, Zacchi, Christian  \&
  Schaffner-Bielich}{Hornick et~al.}{2018}]{PhysRevC.98.065804}
Hornick N.,  Tolos L.,  Zacchi A.,  Christian J.-E.,   Schaffner-Bielich J.,
  2018, \mn@doi [Phys. Rev. C] {10.1103/PhysRevC.98.065804}, 98, 065804

\bibitem[\protect\citeauthoryear{Horowitz \& Piekarewicz}{Horowitz \&
  Piekarewicz}{2001}]{PhysRevLett.86.5647}
Horowitz C.~J.,  Piekarewicz J.,  2001, \mn@doi [Phys. Rev. Lett.]
  {10.1103/PhysRevLett.86.5647}, 86, 5647

\bibitem[\protect\citeauthoryear{{Janka}, {Zwerger}  \& {Moenchmeyer}}{{Janka}
  et~al.}{1993}]{gth2}
{Janka} H.~T.,  {Zwerger} T.,   {Moenchmeyer} R.,  1993, A\&A, \href
  {https://ui.adsabs.harvard.edu/abs/1993A&A...268..360J} {268, 360}

\bibitem[\protect\citeauthoryear{Jeong, McLerran  \& Sen}{Jeong
  et~al.}{2020}]{PhysRevC.101.035201}
Jeong K.~S.,  McLerran L.,   Sen S.,  2020, \mn@doi [Phys. Rev. C]
  {10.1103/PhysRevC.101.035201}, 101, 035201

\bibitem[\protect\citeauthoryear{Keek, Heger  \& in~'t Zand}{Keek
  et~al.}{2012}]{Keek_2012}
Keek L.,  Heger A.,   in~'t Zand J. J.~M.,  2012, \mn@doi [The Astrophysical
  Journal] {10.1088/0004-637X/752/2/150}, 752, 150

\bibitem[\protect\citeauthoryear{Knorren, Prakash  \& Ellis}{Knorren
  et~al.}{1995}]{PhysRevC.52.3470}
Knorren R.,  Prakash M.,   Ellis P.~J.,  1995, \mn@doi [Phys. Rev. C]
  {10.1103/PhysRevC.52.3470}, 52, 3470

\bibitem[\protect\citeauthoryear{Koliogiannis, Kanakis-Pegios  \&
  Moustakidis}{Koliogiannis et~al.}{2021}]{foundations1020017}
Koliogiannis P.~S.,  Kanakis-Pegios A.,   Moustakidis C.~C.,  2021, \mn@doi
  [Foundations] {10.3390/foundations1020017}, 1, 217

\bibitem[\protect\citeauthoryear{Kubis \& Kutschera}{Kubis \&
  Kutschera}{1997}]{KUBIS1997191}
Kubis S.,  Kutschera M.,  1997, \mn@doi [Physics Letters B]
  {https://doi.org/10.1016/S0370-2693(97)00306-7}, 399, 191

\bibitem[\protect\citeauthoryear{Kumar, Biswal  \& Patra}{Kumar
  et~al.}{2017a}]{PhysRevC.95.015801}
Kumar B.,  Biswal S.~K.,   Patra S.~K.,  2017a, \mn@doi [Phys. Rev. C]
  {10.1103/PhysRevC.95.015801}, 95, 015801

\bibitem[\protect\citeauthoryear{Kumar, Singh, Agrawal  \& Patra}{Kumar
  et~al.}{2017b}]{KUMAR2017197}
Kumar B.,  Singh S.,  Agrawal B.,   Patra S.,  2017b, \mn@doi [Nuclear Physics
  A] {https://doi.org/10.1016/j.nuclphysa.2017.07.001}, 966, 197

\bibitem[\protect\citeauthoryear{Kumar, Patra  \& Agrawal}{Kumar
  et~al.}{2018}]{PhysRevC.97.045806}
Kumar B.,  Patra S.~K.,   Agrawal B.~K.,  2018, \mn@doi [Phys. Rev. C]
  {10.1103/PhysRevC.97.045806}, 97, 045806

\bibitem[\protect\citeauthoryear{Kumar, Das, Biswal, Kumar  \& Patra}{Kumar
  et~al.}{2020}]{Kumar2020}
Kumar A.,  Das H.~C.,  Biswal S.~K.,  Kumar B.,   Patra S.~K.,  2020, \mn@doi
  [The European Physical Journal C] {10.1140/epjc/s10052-020-8353-4}, 80, 775

\bibitem[\protect\citeauthoryear{Kutta}{Kutta}{1901}]{rk2}
Kutta W.,  1901, Z. Math. Phys., 46, 435

\bibitem[\protect\citeauthoryear{{Kuulkers} et~al.,}{{Kuulkers}
  et~al.}{2002}]{2002A&A...382..503K}
{Kuulkers} E.,  et~al., 2002, \mn@doi [\aap] {10.1051/0004-6361:20011654},
  \href {https://ui.adsabs.harvard.edu/abs/2002A&A...382..503K} {382, 503}

\bibitem[\protect\citeauthoryear{Lalazissis, K\"onig  \& Ring}{Lalazissis
  et~al.}{1997}]{PhysRevC.55.540}
Lalazissis G.~A.,  K\"onig J.,   Ring P.,  1997, \mn@doi [Phys. Rev. C]
  {10.1103/PhysRevC.55.540}, 55, 540

\bibitem[\protect\citeauthoryear{Lalazissis, Raman  \& Ring}{Lalazissis
  et~al.}{1999}]{LALAZISSIS19991}
Lalazissis G.,  Raman S.,   Ring P.,  1999, \mn@doi [Atomic Data and Nuclear
  Data Tables] {https://doi.org/10.1006/adnd.1998.0795}, 71, 1

\bibitem[\protect\citeauthoryear{Lalazissis, Karatzikos, Fossion, Arteaga,
  Afanasjev  \& Ring}{Lalazissis et~al.}{2009}]{LALAZISSIS200936}
Lalazissis G.,  Karatzikos S.,  Fossion R.,  Arteaga D.~P.,  Afanasjev A.,
  Ring P.,  2009, \mn@doi [Physics Letters B]
  {https://doi.org/10.1016/j.physletb.2008.11.070}, 671, 36

\bibitem[\protect\citeauthoryear{Lattimer}{Lattimer}{2012}]{doi:10.1146/annurev-nucl-102711-095018}
Lattimer J.~M.,  2012, \mn@doi [Annual Review of Nuclear and Particle Science]
  {10.1146/annurev-nucl-102711-095018}, 62, 485

\bibitem[\protect\citeauthoryear{Lattimer}{Lattimer}{2021}]{doi:10.1146/annurev-nucl-102419-124827}
Lattimer J.,  2021, \mn@doi [Annual Review of Nuclear and Particle Science]
  {10.1146/annurev-nucl-102419-124827}, 71, 433

\bibitem[\protect\citeauthoryear{Leung, Chu  \& Lin}{Leung
  et~al.}{2022}]{PhysRevD.105.123010}
Leung K.-L.,  Chu M.-c.,   Lin L.-M.,  2022, \mn@doi [Phys. Rev. D]
  {10.1103/PhysRevD.105.123010}, 105, 123010

\bibitem[\protect\citeauthoryear{Linares, van~der Klis, Altamirano  \&
  Markwardt}{Linares et~al.}{2005}]{Linares_2005}
Linares M.,  van~der Klis M.,  Altamirano D.,   Markwardt C.~B.,  2005, \mn@doi
  [The Astrophysical Journal] {10.1086/497025}, 634, 1250

\bibitem[\protect\citeauthoryear{Liu, Greco, Baran, Colonna  \& Di~Toro}{Liu
  et~al.}{2002}]{PhysRevC.65.045201}
Liu B.,  Greco V.,  Baran V.,  Colonna M.,   Di~Toro M.,  2002, \mn@doi [Phys.
  Rev. C] {10.1103/PhysRevC.65.045201}, 65, 045201

\bibitem[\protect\citeauthoryear{Löffler et~al.,}{Löffler et~al.}{2012}]{ET2}
Löffler F.,  et~al., 2012, \mn@doi [Classical and Quantum Gravity]
  {10.1088/0264-9381/29/11/115001}, 29, 115001

\bibitem[\protect\citeauthoryear{Mata~Carrizal, Valbuena~Ordóñez,
  Garza~Aguirre, Betancourt~Sotomayor  \& Morones~Ibarra}{Mata~Carrizal
  et~al.}{2022}]{universe8050264}
Mata~Carrizal N.~B.,  Valbuena~Ordóñez E.,  Garza~Aguirre A.~J.,
  Betancourt~Sotomayor F.~J.,   Morones~Ibarra J.~R.,  2022, \mn@doi [Universe]
  {10.3390/universe8050264}, 8

\bibitem[\protect\citeauthoryear{McLerran \& Reddy}{McLerran \&
  Reddy}{2019}]{PhysRevLett.122.122701}
McLerran L.,  Reddy S.,  2019, \mn@doi [Phys. Rev. Lett.]
  {10.1103/PhysRevLett.122.122701}, 122, 122701

\bibitem[\protect\citeauthoryear{Menezes \& Provid\^encia}{Menezes \&
  Provid\^encia}{2004}]{PhysRevC.70.058801}
Menezes D.~P.,  Provid\^encia C.,  2004, \mn@doi [Phys. Rev. C]
  {10.1103/PhysRevC.70.058801}, 70, 058801

\bibitem[\protect\citeauthoryear{Miao, Jiang, Li  \& Chen}{Miao
  et~al.}{2021}]{Miao_2021}
Miao Z.,  Jiang J.-L.,  Li A.,   Chen L.-W.,  2021, \mn@doi [The Astrophysical
  Journal Letters] {10.3847/2041-8213/ac194d}, 917, L22

\bibitem[\protect\citeauthoryear{{Miller} et~al.,}{{Miller}
  et~al.}{2021}]{2021ApJ...918L..28M}
{Miller} M.~C.,  et~al., 2021, \mn@doi [\apjl] {10.3847/2041-8213/ab960f},
  \href {https://ui.adsabs.harvard.edu/abs/2021ApJ...918L..28M} {896, L44}

\bibitem[\protect\citeauthoryear{Müller \& Serot}{Müller \&
  Serot}{1996}]{MULLER1996508}
Müller H.,  Serot B.~D.,  1996, \mn@doi [Nuclear Physics A]
  {https://doi.org/10.1016/0375-9474(96)00187-X}, 606, 508

\bibitem[\protect\citeauthoryear{Nakamura, Oohara  \& Kojima}{Nakamura
  et~al.}{1987}]{bssn1}
Nakamura T.,  Oohara K.,   Kojima Y.,  1987, \mn@doi [Progress of Theoretical
  Physics Supplement] {10.1143/PTPS.90.1}, 90, 1

\bibitem[\protect\citeauthoryear{Newman \& Penrose}{Newman \&
  Penrose}{1963}]{np}
Newman E.,  Penrose R.,  1963, \mn@doi [Journal of Mathematical Physics]
  {10.1063/1.1704025}, 4, 998

\bibitem[\protect\citeauthoryear{Noble, Gammie, McKinney  \& Zanna}{Noble
  et~al.}{2006}]{p2c1}
Noble S.~C.,  Gammie C.~F.,  McKinney J.~C.,   Zanna L.~D.,  2006, \mn@doi [The
  Astrophysical Journal] {10.1086/500349}, 641, 626

\bibitem[\protect\citeauthoryear{Noble, Krolik  \& Hawley}{Noble
  et~al.}{2009}]{p2c2}
Noble S.~C.,  Krolik J.~H.,   Hawley J.~F.,  2009, \mn@doi [The Astrophysical
  Journal] {10.1088/0004-637x/692/1/411}, 692, 411

\bibitem[\protect\citeauthoryear{Oppenheimer \& Volkoff}{Oppenheimer \&
  Volkoff}{1939}]{PhysRev.55.374}
Oppenheimer J.~R.,  Volkoff G.~M.,  1939, \mn@doi [Phys. Rev.]
  {10.1103/PhysRev.55.374}, 55, 374

\bibitem[\protect\citeauthoryear{Parmar, Das, Kumar, Sharma  \& Patra}{Parmar
  et~al.}{2022}]{PhysRevD.105.043017}
Parmar V.,  Das H.~C.,  Kumar A.,  Sharma M.~K.,   Patra S.~K.,  2022, \mn@doi
  [Phys. Rev. D] {10.1103/PhysRevD.105.043017}, 105, 043017

\bibitem[\protect\citeauthoryear{Pawar, Kalamkar, Altamirano, Linares, Shanthi,
  Strohmayer, Bhattacharya  \& Klis}{Pawar et~al.}{2013}]{10.1093/mnras/stt919}
Pawar D.~D.,  Kalamkar M.,  Altamirano D.,  Linares M.,  Shanthi K.,
  Strohmayer T.,  Bhattacharya D.,   Klis M. v.~d.,  2013, \mn@doi [Monthly
  Notices of the Royal Astronomical Society] {10.1093/mnras/stt919}, 433, 2436

\bibitem[\protect\citeauthoryear{Prasad \& Mallick}{Prasad \&
  Mallick}{2018}]{Prasad2018-ck}
Prasad R.,  Mallick R.,  2018, Astrophys. J., 859, 57

\bibitem[\protect\citeauthoryear{Rashdan}{Rashdan}{2001}]{PhysRevC.63.044303}
Rashdan M.,  2001, \mn@doi [Phys. Rev. C] {10.1103/PhysRevC.63.044303}, 63,
  044303

\bibitem[\protect\citeauthoryear{Read, Lackey, Owen  \& Friedman}{Read
  et~al.}{2009}]{ppeos}
Read J.~S.,  Lackey B.~D.,  Owen B.~J.,   Friedman J.~L.,  2009, \mn@doi [Phys.
  Rev. D] {10.1103/PhysRevD.79.124032}, 79, 124032

\bibitem[\protect\citeauthoryear{Reinhard}{Reinhard}{1989}]{Reinhard_1989}
Reinhard P.~G.,  1989, \mn@doi [Reports on Progress in Physics]
  {10.1088/0034-4885/52/4/002}, 52, 439

\bibitem[\protect\citeauthoryear{Reisswig, Ott, Sperhake  \&
  Schnetter}{Reisswig et~al.}{2011}]{mclachlan1}
Reisswig C.,  Ott C.~D.,  Sperhake U.,   Schnetter E.,  2011, \mn@doi [Phys.
  Rev. D] {10.1103/PhysRevD.83.064008}, 83, 064008

\bibitem[\protect\citeauthoryear{Rezzolla \& Takami}{Rezzolla \&
  Takami}{2016}]{gw_ana}
Rezzolla L.,  Takami K.,  2016, \mn@doi [Phys. Rev. D]
  {10.1103/PhysRevD.93.124051}, 93, 124051

\bibitem[\protect\citeauthoryear{Rezzolla \& Zanotti}{Rezzolla \&
  Zanotti}{2013}]{adm13}
Rezzolla L.,  Zanotti O.,  2013, Relativistic Hydrodynamics.
Oxford University Press, \mn@doi{10.1093/acprof:oso/9780198528906.001.0001}

\bibitem[\protect\citeauthoryear{Ridolfi et~al.,}{Ridolfi
  et~al.}{2021}]{10.1093/mnras/stab790}
Ridolfi A.,  et~al., 2021, \mn@doi [Monthly Notices of the Royal Astronomical
  Society] {10.1093/mnras/stab790}, 504, 1407

\bibitem[\protect\citeauthoryear{Riley et~al.,}{Riley
  et~al.}{2021}]{Riley_2021}
Riley T.~E.,  et~al., 2021, \mn@doi [The Astrophysical Journal Letters]
  {10.3847/2041-8213/ac0a81}, 918, L27

\bibitem[\protect\citeauthoryear{Roca-Maza, Vi\~nas, Centelles, Ring  \&
  Schuck}{Roca-Maza et~al.}{2011}]{PhysRevC.84.054309}
Roca-Maza X.,  Vi\~nas X.,  Centelles M.,  Ring P.,   Schuck P.,  2011, \mn@doi
  [Phys. Rev. C] {10.1103/PhysRevC.84.054309}, 84, 054309

\bibitem[\protect\citeauthoryear{Romani, Kandel, Filippenko, Brink  \&
  Zheng}{Romani et~al.}{2022}]{Romani_2022}
Romani R.~W.,  Kandel D.,  Filippenko A.~V.,  Brink T.~G.,   Zheng W.,  2022,
  \mn@doi [The Astrophysical Journal Letters] {10.3847/2041-8213/ac8007}, 934,
  L17

\bibitem[\protect\citeauthoryear{Runge}{Runge}{1895}]{rk1}
Runge C.,  1895, \mn@doi [Mathematische Annalen] {10.1007/BF01446807}, 46, 167

\bibitem[\protect\citeauthoryear{Russotto, Cozma, De~Filippo, Le~F{\`e}vre,
  Leifels  \& {\L}ukasik}{Russotto et~al.}{2023}]{Russotto2023}
Russotto P.,  Cozma M.~D.,  De~Filippo E.,  Le~F{\`e}vre A.,  Leifels Y.,
  {\L}ukasik J.,  2023, \mn@doi [La Rivista del Nuovo Cimento]
  {10.1007/s40766-023-00039-4}, 46, 1

\bibitem[\protect\citeauthoryear{Schnetter, Hawley  \& Hawke}{Schnetter
  et~al.}{2004}]{amr}
Schnetter E.,  Hawley S.~H.,   Hawke I.,  2004, \mn@doi [Classical and Quantum
  Gravity] {10.1088/0264-9381/21/6/014}, 21, 1465

\bibitem[\protect\citeauthoryear{Schnetter, Diener, Dorband  \&
  Tiglio}{Schnetter et~al.}{2006}]{carpet1}
Schnetter E.,  Diener P.,  Dorband E.~N.,   Tiglio M.,  2006, \mn@doi
  [Classical and Quantum Gravity] {10.1088/0264-9381/23/16/s14}, 23, S553

\bibitem[\protect\citeauthoryear{Serot}{Serot}{1979}]{SEROT1979146}
Serot B.~D.,  1979, \mn@doi [Physics Letters B]
  {https://doi.org/10.1016/0370-2693(79)90804-9}, 86, 146

\bibitem[\protect\citeauthoryear{Serot \& Walecka}{Serot \&
  Walecka}{1986}]{Serot:1984ey}
Serot B.~D.,  Walecka J.~D.,  1986, Adv. Nucl. Phys., 16, 1

\bibitem[\protect\citeauthoryear{Serot \& Walecka}{Serot \&
  Walecka}{1997}]{doi:10.1142/S0218301397000299}
Serot B.~D.,  Walecka J.~D.,  1997, \mn@doi [International Journal of Modern
  Physics E] {10.1142/S0218301397000299}, 06, 515

\bibitem[\protect\citeauthoryear{Shibata \& Nakamura}{Shibata \&
  Nakamura}{1995}]{bssn2}
Shibata M.,  Nakamura T.,  1995, \mn@doi [Phys. Rev. D]
  {10.1103/PhysRevD.52.5428}, 52, 5428

\bibitem[\protect\citeauthoryear{Sotani \& Kokkotas}{Sotani \&
  Kokkotas}{2017}]{PhysRevD.95.044032}
Sotani H.,  Kokkotas K.~D.,  2017, \mn@doi [Phys. Rev. D]
  {10.1103/PhysRevD.95.044032}, 95, 044032

\bibitem[\protect\citeauthoryear{Sulaksono \& Mart}{Sulaksono \&
  Mart}{2006}]{PhysRevC.74.045806}
Sulaksono A.,  Mart T.,  2006, \mn@doi [Phys. Rev. C]
  {10.1103/PhysRevC.74.045806}, 74, 045806

\bibitem[\protect\citeauthoryear{Takami, Rezzolla  \& Baiotti}{Takami
  et~al.}{2015}]{gth}
Takami K.,  Rezzolla L.,   Baiotti L.,  2015, \mn@doi [Phys. Rev. D]
  {10.1103/PhysRevD.91.064001}, 91, 064001

\bibitem[\protect\citeauthoryear{Tews, Carlson, Gandolfi  \& Reddy}{Tews
  et~al.}{2018}]{Tews_2018}
Tews I.,  Carlson J.,  Gandolfi S.,   Reddy S.,  2018, \mn@doi [The
  Astrophysical Journal] {10.3847/1538-4357/aac267}, 860, 149

\bibitem[\protect\citeauthoryear{Thornburg}{Thornburg}{2003}]{ETextra1}
Thornburg J.,  2003, \mn@doi [Classical and Quantum Gravity]
  {10.1088/0264-9381/21/2/026}, 21, 743

\bibitem[\protect\citeauthoryear{Todd-Rutel \& Piekarewicz}{Todd-Rutel \&
  Piekarewicz}{2005}]{PhysRevLett.95.122501}
Todd-Rutel B.~G.,  Piekarewicz J.,  2005, \mn@doi [Phys. Rev. Lett.]
  {10.1103/PhysRevLett.95.122501}, 95, 122501

\bibitem[\protect\citeauthoryear{Tolman}{Tolman}{1939}]{PhysRev.55.364}
Tolman R.~C.,  1939, \mn@doi [Phys. Rev.] {10.1103/PhysRev.55.364}, 55, 364

\bibitem[\protect\citeauthoryear{Tyul'bashev, Tyul'bashev, Oreshko  \&
  Logvinenko}{Tyul'bashev et~al.}{2016}]{Tyulbashev2016}
Tyul'bashev S.~A.,  Tyul'bashev V.~S.,  Oreshko V.~V.,   Logvinenko S.~V.,
  2016, \mn@doi [Astronomy Reports] {10.1134/S1063772916020128}, 60, 220

\bibitem[\protect\citeauthoryear{{Vahdat}, {Posselt}, {Santangelo}  \&
  {Pavlov}}{{Vahdat} et~al.}{2022}]{2022A&A...658A..95V}
{Vahdat} A.,  {Posselt} B.,  {Santangelo} A.,   {Pavlov} G.~G.,  2022, \mn@doi
  [\aap] {10.1051/0004-6361/202141795}, \href
  {https://ui.adsabs.harvard.edu/abs/2022A&A...658A..95V} {658, A95}

\bibitem[\protect\citeauthoryear{Walecka}{Walecka}{1974}]{WALECKA1974491}
Walecka J.,  1974, \mn@doi [Annals of Physics]
  {https://doi.org/10.1016/0003-4916(74)90208-5}, 83, 491

\bibitem[\protect\citeauthoryear{Weissenborn, Chatterjee  \&
  Schaffner-Bielich}{Weissenborn et~al.}{2012}]{PhysRevC.85.065802}
Weissenborn S.,  Chatterjee D.,   Schaffner-Bielich J.,  2012, \mn@doi [Phys.
  Rev. C] {10.1103/PhysRevC.85.065802}, 85, 065802

\bibitem[\protect\citeauthoryear{{Wijnands}, {van der Klis}, {Homan},
  {Chakrabarty}, {Markwardt}  \& {Morgan}}{{Wijnands}
  et~al.}{2003}]{2003HEAD....7.1702W}
{Wijnands} R.,  {van der Klis} M.,  {Homan} J.,  {Chakrabarty} D.,  {Markwardt}
  C.~B.,   {Morgan} E.~H.,  2003, in AAS/High Energy Astrophysics Division \#7.
  p. 17.02

\bibitem[\protect\citeauthoryear{York}{York}{1999}]{lorene3}
York J.~W.,  1999, \mn@doi [Phys. Rev. Lett.] {10.1103/PhysRevLett.82.1350},
  82, 1350

\bibitem[\protect\citeauthoryear{Zhang, Hu  \& Liu}{Zhang
  et~al.}{2018}]{PhysRevC.97.015805}
Zhang Y.,  Hu J.,   Liu P.,  2018, \mn@doi [Phys. Rev. C]
  {10.1103/PhysRevC.97.015805}, 97, 015805

\bibitem[\protect\citeauthoryear{Zhao \& Lattimer}{Zhao \&
  Lattimer}{2020}]{PhysRevD.102.023021}
Zhao T.,  Lattimer J.~M.,  2020, \mn@doi [Phys. Rev. D]
  {10.1103/PhysRevD.102.023021}, 102, 023021

\bibitem[\protect\citeauthoryear{Zhu, Li  \& Rezzolla}{Zhu
  et~al.}{2020}]{PhysRevD.102.084058}
Zhu Z.,  Li A.,   Rezzolla L.,  2020, \mn@doi [Phys. Rev. D]
  {10.1103/PhysRevD.102.084058}, 102, 084058

\bibitem[\protect\citeauthoryear{{in't Zand}, {Kries}, {Palmer}  \&
  {Degenaar}}{{in't Zand} et~al.}{2019}]{2019A&A...621A..53I}
{in't Zand} J.~J.~M.,  {Kries} M.~J.~W.,  {Palmer} D.~M.,   {Degenaar} N.,
  2019, \mn@doi [\aap] {10.1051/0004-6361/201834270}, \href
  {https://ui.adsabs.harvard.edu/abs/2019A&A...621A..53I} {621, A53}

\makeatother
\end{thebibliography}
\bibliographystyle{mnras}
\end{document}